\documentclass{emulateapj}
\usepackage{amsmath}
\usepackage{graphicx}
\usepackage{natbib}
\usepackage[scaled]{helvet}
\usepackage{epsfig}
\usepackage{url}
\bibpunct{(}{)}{;}{a}{}{,}
\usepackage{textcomp}
\usepackage{mathpazo}
\usepackage{amsmath}
\usepackage{rotating}

\usepackage{color}
\usepackage[normalem]{ulem}

\newcommand{\msun}{\ensuremath{\,M_\Sun}}

\begin{document}

\title{Identification of Young Stellar Variables with KELT for \emph{K2} I: Taurus Dippers and Rotators}
\author{Joseph E. Rodriguez$^1$, 
Megan Ansdell$^{2}$, 
Ryan J. Oelkers$^{3}$, 
Phillip A. Cargile$^{1}$, 
Eric Gaidos$^{4,5}$, 
Ann Marie Cody$^{6}$, 
Daniel J. Stevens$^{7}$,
Garrett Somers$^{3}$,
David James$^{8}$,
Thomas G. Beatty$^{9,10}$, 
Robert J. Siverd$^{11}$,
Michael B. Lund$^{3}$,
Rudolf B. Kuhn$^{12}$,
B. Scott Gaudi$^{7}$,
Joshua Pepper$^{13}$,
Keivan G. Stassun$^{3,14}$
}

\affil{$^{1}$Harvard-Smithsonian Center for Astrophysics, 60 Garden St, Cambridge, MA 02138, USA}
\affil{$^{2}$Institute for Astronomy, University of Hawai`i at Manoa, Honolulu, HI 96822, USA}
\affil{$^{3}$Department of Physics and Astronomy, Vanderbilt University, 6301 Stevenson Center, Nashville, TN 37235, USA}
\affil{$^{4}$Department of Geology and Geophysics, University of Hawai`i at Manoa, Honolulu, HI 96822}
\affil{$^{5}$Fulbright Fellow, Institute for Astrophysics, University of Vienna, Vienna AT-1180, Austria}
\affil{$^{6}$NASA Ames Research Center, Mountain View, CA 94035, USA}
\affil{$^{7}$Department of Astronomy, The Ohio State University, Columbus, OH 43210, USA}
\affil{$^{8}$Astronomy Department, University of Washington, Box 351580, Seattle, WA 98195, USA}
\affil{$^{9}$Department of Astronomy \& Astrophysics, The Pennsylvania State University, 525 Davey Lab, University Park, PA 16802}
\affil{$^{10}$Center for Exoplanets and Habitable Worlds, The Pennsylvania State University, 525 Davey Lab, University Park, PA 16802}
\affil{$^{11}$Las Cumbres Observatory Global Telescope Network, 6740 Cortona Dr., Suite 102, Santa Barbara, CA 93117, USA}
\affil{$^{12}$South African Astronomical Observatory, PO Box 9, Observatory 7935, South Africa}
\affil{$^{13}$Department of Physics, Lehigh University, 16 Memorial Drive East, Bethlehem, PA 18015, USA}
\affil{$^{14}$Department of Physics, Fisk University, 1000 17th Avenue North, Nashville, TN 37208, USA}

%\affil{$^{12}$NASA Goddard Space Flight Center, Greenbelt, MD 20771, USA}

\shorttitle{Taurus Dippers}

\begin{abstract}
One of the most well-studied young stellar associations, Taurus-Auriga, was observed by the extended Kepler mission, {\it K2}, in the spring of 2017. {\it K2} Campaign 13 (C13) is a unique opportunity to study many stars in this young association at high photometric precision and cadence. Using observations from the Kilodegree Extremely Little Telescope (KELT) survey, we identify ``dippers", aperiodic and periodic variables among {\it K2} C13 target stars. This release of the KELT data  (lightcurve data in e-tables) provides the community with long-time baseline observations to assist in the understanding of the more exotic variables in the association. Transient-like phenomena on timescales of months to years are known characteristics in the light curves of young stellar objects, making contextual pre- and post-{\it K2} observations critical to understanding their underlying processes. We are providing a comprehensive set of the KELT light curves for known Taurus-Auriga stars in \emph{K2} C13. The combined data sets from \emph{K2} and KELT should permit a broad array of investigations related to star formation, stellar variability, and protoplanetary environments.

\end{abstract}

\keywords{circumstellar matter, protoplanetary disks, stars: pre-main sequence, stars: variables: T Tauri}
\shortauthors{Rodriguez et al.}

\section{Introduction}
The study of the photometric variability of young stellar objects (YSOs) goes back to the pioneering work of \citet{Joy:1949}, who first associated variability in T~Tauri stars (TTSs) with the star formation process. Subsequent decades of photometric and spectroscopic investigations have painted a rich picture of the nature of TTSs and their variability processes, which can include coherent periodic variations arising from surface star spots (akin to sunspots) as well as quasi-coherent and aperiodic variability arising from accretion and outflows, likely arising from magnetospheric interactions between the star and its protoplanetary disk \citep[for an early comprehensive review, see][and references therein]{Bertout:1989}. Additionally, TTSs have shown variability, both short and long in duration, caused by circumstellar extinction from dust or part of the surrounding circumstellar disk \citep{Herbst:1994, Bouvier:1999, Bouvier:2013}.
% EG: mention occultation by circumstellar dust/disks as another source of variability. JR Done

Therefore, studying the variability of TTSs has been key to developing insight into the early stages of star formation, stellar angular momentum evolution, the physics of magnetic star-disk interaction, and the nature of the protoplanetary gas and dust that lead to planet formation. With the success of long-time baseline surveys (All-Sky Automated Survey, Catalina Real-Time Transient Survey, and Palomar Transient Factory \citealp{Pojamanski:1997, Drake:2009, Law:2009}), the number of long-time baseline photometric observations of young stellar associations has significantly increased. From these long-time baseline, high-cadence observations, we are now able to study these systems in much greater detail.

\begin{figure*}[!ht]
  \centering
  \includegraphics[width=0.75\linewidth, angle=-90, trim = 0 0 0 0 ]{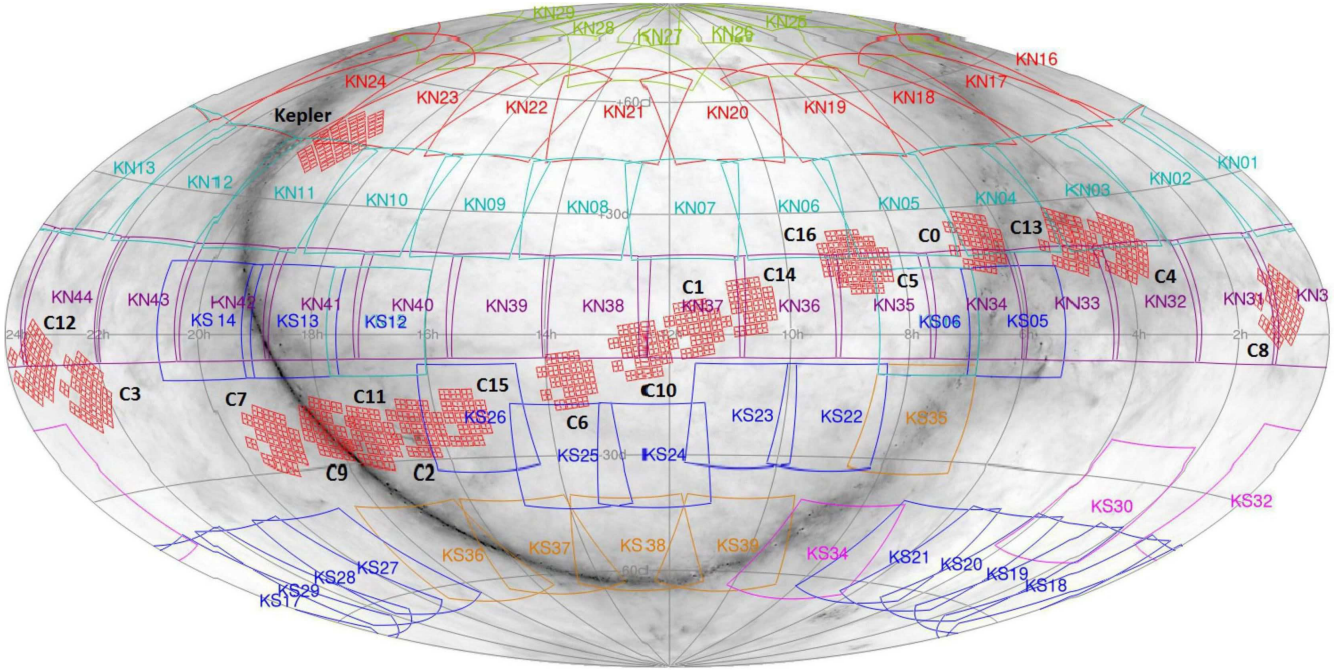}
  \vspace{-0.8in}
  \caption{Map in celestial coordinates showing the locations of all of the KELT-North and KELT-South fields. The colors of each field indicates when the field was added to the normal observing scheme (Cyan: 2006; Blue: 2010; Red: 2012; Orange: 2013; Yellow: 2014). The purple equatorial fields were added in 2015 and only observed by KELT-North. The purple fields labeled KS34 and KS32 are the original KELT-South commissioning fields that were observed in 2009 and then added to the normal observing schedule in late-2011. The locations of the \emph{Kepler} and \emph{K2} fields are indicated by the red outline of each Kepler detector. This figure was created using the Montage Image Mosaic Engine \citep{Berriman:2017}.}
 \label{fig:fields}
\end{figure*}

As a result of observed spectroscopic signatures, TTSs are typically split into two sub-groups, weak-lined T~Tauri stars (WTTSs) and classical T~Tauri stars (CTTSs). One key characteristic distinguishing the two sub-groups is the level of Balmer series H$\alpha$ emission observed in their optical spectra. CTTSs have H$\alpha$ equivalent widths (EWs) $>$5-10\r{A}, whereas WTTSs are more sedate (EW$_{H\alpha}$ $\ll$  5\r{A}) in their emission \citep{Martin:1998,Barrado:2003}. However, this cutoff between CTTSs and WTTSs changes as a function of spectral type, and is closer to $>$10\r{A} for M dwarfs \citep{White:2003}. 

The variability for these two types of YSOs can also be quite different. WTTSs tend to display relatively stable sinusoidal variability believed to be caused by surface star spots rotating in and out view \citep{Stassun:1999, Grankin:2008,Rodriguez-Ledesma:2009, Frasca:2009}. In contrast, CTTSs display a wide variety of photometric variability which has been observed to be periodic, semi-periodic, or non-periodic \citep{Herbst:1994}. The semi-periodic and periodic variability in CTTSs has been attributed to circumstellar extinction \citep{Herbst:1994, Chelli:1999, Alencar:2010}, occultations by an orbiting body in Keplerian motion \citep{Mamajek:2012,Bouvier:2013, Rodriguez:2013, Rodriguez:2016A}, or accretion --- either in a disk or onto the star \citep{Bertout:1988, Bertout:1989,Stassun:1999, Carpenter:2002, Scholz:2009, Bouvier:2007}. Additionally, there is a new appreciation for a broad class of YSO variables that involve obscuration by disk material, referred to as ``disk eclipsing'' systems \citep[see, e.g.,][and references therein]{Herbst:2010, Plavchan:2013, Rodriguez:2015, Rodriguez:2016C}, whereby the star is occulted by features in its protoplanetary disk that may represent advanced stages of planet formation and that could provide a probe of protoplanetary material.

The Taurus-Aurigae association is the nearest (140~pc) and arguably the best-studied association of YSOs, including the eponymous T Tauri stars, and their variability \citep[see, e.g.,][and references therein as well as above]{Bertout:1989}. However, studying the variability of Taurus stars is challenging as the objects are spread over hundreds of square degrees. The advent of continuous, high-cadence, wide-field surveys has transformed our ability to comprehensively and systematically study these systems.  The re-purposed {\it Kepler} mission, {\it K2} \citep{Howell:2014}, has already supplied the community with high-precision photometry of the young stellar associations Upper Sco and $\rho$~Oph \citep{Ripepi:2015, Ansdell:2016a, Scaringi:2016, Cody:2017, Stauffer:2017}. Unfortunately, observations from {\it K2} are only taken in one band-pass, and the data are not available until months after they are taken. Thus, relying on those data alone limits our ability to identify and characterize variable stars of interest. Additionally, it is difficult to complement \emph{K2} with simultaneous multi-wavelength observations. Multi-band photometric and spectroscopic observations can provide crucial information for identifying the sub-class of YSO variability. Therefore, to enable simultaneous {\it K2}-ground-based observations of variable young stars, we used photometric observations from the Kilodegree Extremely Little Telescope (KELT) survey to identify dippers and periodic signals of 56 YSOs in the Taurus-Auriga association. Of these, fifty of the stars in our sample were observed by \emph{K2} during Campaign 13. YSOs can display a wide range of variability, and the observing baseline, sky coverage, and precision of the KELT observations provide the unique ability to study these types of variability; from the short-duration events seen in dippers to the disk eclipsing systems that can dim for many months to years. We are releasing the results of our analysis and the corresponding data to extend the photometric time baseline of these targets, and enable complementary and simultaneous ground-based observations.   

In this paper, we present time-series photometry from the KELT survey of known YSOs in the Taurus-Auriga association that were observed in \emph{K2} Campaign 13 (C13). Our target selection for this study is described in \S2. The photometric observations are presented in \S3. We present a robust estimate of the stellar parameters for each star in our sample and our methodology is shown in \S4. We describe our technique for characterizing variability and periodicity in \S5, overview the different identified variables in \S6, and summarize our results in \S7.

\begin{figure*}[!ht]
  \centering
  \includegraphics[width=0.8\textwidth]{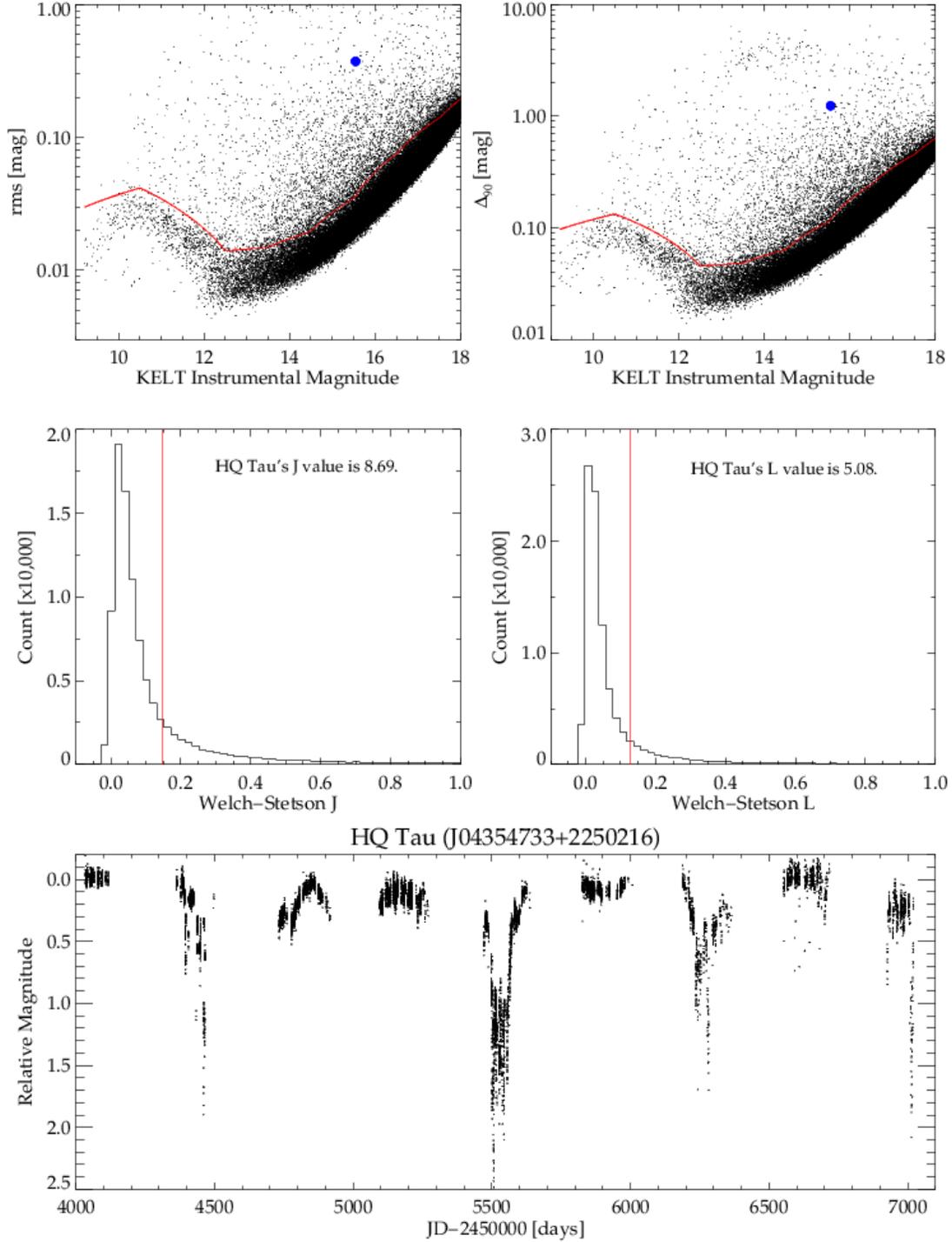}
  \caption{The variability metrics used to select \emph{K2} Taurus members as possible variable objects using the ensemble light curves found in KELT-North field 03. Stars lying above the red line in the top panels and to the right of the red line in the middle panels are expected to show variability due to astrophysical phenomena and not systematics. The cutoffs described below were calculated after an iterative 2.5$\sigma$ clipping of the mean of each distribution to remove possible outliers. \textit{Top Row}: The \textit{rms} (left) and $\Delta_{90}$ (right) statistics and their respective, independent $+2\sigma$ quartiles shown as a red line. \textit{Middle Row}: The Welch-Stetson \textit{J} \& \textit{L} statistics and their respective $+3\sigma_{J}$ \& $+5\sigma_{L}$ cuts, calculated using objects with $J,L<3$ and a 2.5$\sigma$ iterative clipping process. We find the KELT-North field 03 has a distribution of \textit{J} values with a mean value of $0.054\pm0.031$, leading to a cut of 0.147; and a distribution of \textit{L} values with a mean of $0.033\pm0.019$, leading to a cut of 0.128. \textit{Bottom Row}: The light curve of HQ Tau. HQ Tau passed all 4 variability metrics and is identified as statistically variable. The star is shown as a blue dot in the top panels and its \textit{J} \& \textit{L} values are labeled in the middle panels.}
 \label{fig:UXor}
\end{figure*}

\section{Target Selection}
\label{sec:selection}
Taurus members were selected by combining the {\it Spitzer} infrared color-selected catalog of \citet{Luhman:2010}, which was in turn augmented with {\it WISE} data by \citet{Esplin:2014}, with the ultraviolet (UV)-selected catalog of \citet{Gomes:2015}, the Taurus Molecular Cloud sources identified as {\it Rosat} X-ray sources by \citet{Carkner:1997}, and the Taurus stars matched with X-ray sources from the {\it XMM-Newton} survey performed by \citet{Guedel:2007}.   Only sources that had an infrared counterpart in either the 2MASS or {\it WISE} catalogs were retained.  This concatenation produced 794 candidates, of which 295 were found to fall on working \emph{K2} detectors in C13.  

Three stars have proper motions consistent with membership in the Hyades cluster and one star has a {\it Hipparcos} parallax identifying it as a foreground star.  We excluded several other stars that have proper motions that deviate by more than 50 mas~yr$^{-1}$ from an iteratively-calculated median Taurus Molecular Cloud motion of $\mu_{\alpha} = +5$~mas~yr$^{-1}$, $\mu_{\delta} = -16$~mas~yr$^{-1}$, consistent with the motion found by \citet{Frink:1997}.

We estimated {\it Kepler} $K_p$ magnitudes to select those stars sufficiently bright ($K_p < 19$) to detect the 10\% flux variation over a day characteristic of ``dipper" stars.  This corresponds to $\le1$\% photometric precision for {\it K2} over 6.5~hr.  $K_p$ magnitudes were calculated using the most reliable method available, first with APASS or SDSS $g$ and $r$ magnitudes, using USNO-B $B1$ and $R1$ as a last resort. This final catalog contained 207 Taurus molecular cloud (TMC) members (median $K_p = 14.4$) and was then cross-matched to the KELT survey, resulting in 56 matches. However, due to the large pixel scale of the KELT telescopes some of these targets are blended in a single KELT lightcurve (see \S \ref{sec:variability} for how blends were identified). The catalog broad-band magnitudes, kinematics, and estimated $K_p$ magnitudes for these 56 sources are presented in Tables \ref{tab:CatalogueInfo} and \ref{tab:CatalogueInfo2}.  

\section{Observations: KELT}

To identify variable YSOs among the Taurus YSOs prior to the {\it K2} C13 observations (See Figure \ref{fig:fields} for the \emph{K2} and KELT fields), we used observations from the KELT exoplanet transit survey. Designed to discover giant planets transiting bright ($V<11$) host stars, the KELT survey uses two 42 mm telephoto lenses (KELT-North at Winer Observatory in Arizona in the United States and KELT-South at the South African Astronomical Observatory in Sutherland, South Africa). The two telescopes combine to observe over 70\% of the entire sky with a 10-30 minute cadence. Each telescope setup provides a $26^\circ$ x $26^\circ$ field of view and a 23$\arcsec$ pixel scale. The typical photometric error for stars 7$<V<$11 (the target brightness range) is $\sim$1\% but the telescopes also obtain lower precision observations of stars down to $V\sim$14 \citep{Pepper:2007, Pepper:2012}. Each KELT telescope uses a Paramount ME robotic German equatorial mount which requires a meridian flip when crossing from an east to west orientation. The east and west images are reduced and extracted separately. For a detailed description of the KELT observing strategy and data reduction process, see \citet{Siverd:2012} and \citet{Kuhn:2016}. The light curves used in this work are from KELT-North field 03 which is centered at J2000 $\alpha$ = 03h 58m 12.0s, $\delta$ = $+31^{\circ}$ 39$\arcmin$ 56.16$\arcsec$ with the exception of J05080709+2427123 which is in KELT-North field 04 centered at J2000 $\alpha$ = 05h 54m 14.5s, $\delta$ = $+31^{\circ}$ 39$\arcmin$ 56.16$\arcsec$. 

The per-point photometric error for our sample is dependent on the brightness of each target and ranges from 0.009 to 0.15 mag. Both KELT telescopes observe only with a non-standard broad $R$-band filter. Therefore, the reported KELT magnitudes in this paper do not correspond to any standard filter and are only instrumental magnitudes.  All plots showing the KELT light curves are in relative magnitude where the median of the entire light curve has been subtracted off. The electronic tables being published with this paper report the KELT instrumental magnitudes and the corresponding instrumental per point error. The KELT light curves and catalog information for all 56 stars are publicly available as electronic table using a Filtergraph\footnote{\url{https: \\filtergraph.com/kelt\_k2}, This link includes KELT light curves for other papers in this series} portal \citep{Burger:2013}.

We recommend only using KELT data where the relative flux uncertainty is $<$20\% rms.

\begin{figure*}[!ht]
  \centering
  \includegraphics[width=0.99\linewidth, trim = 0 2.3in 0 0]{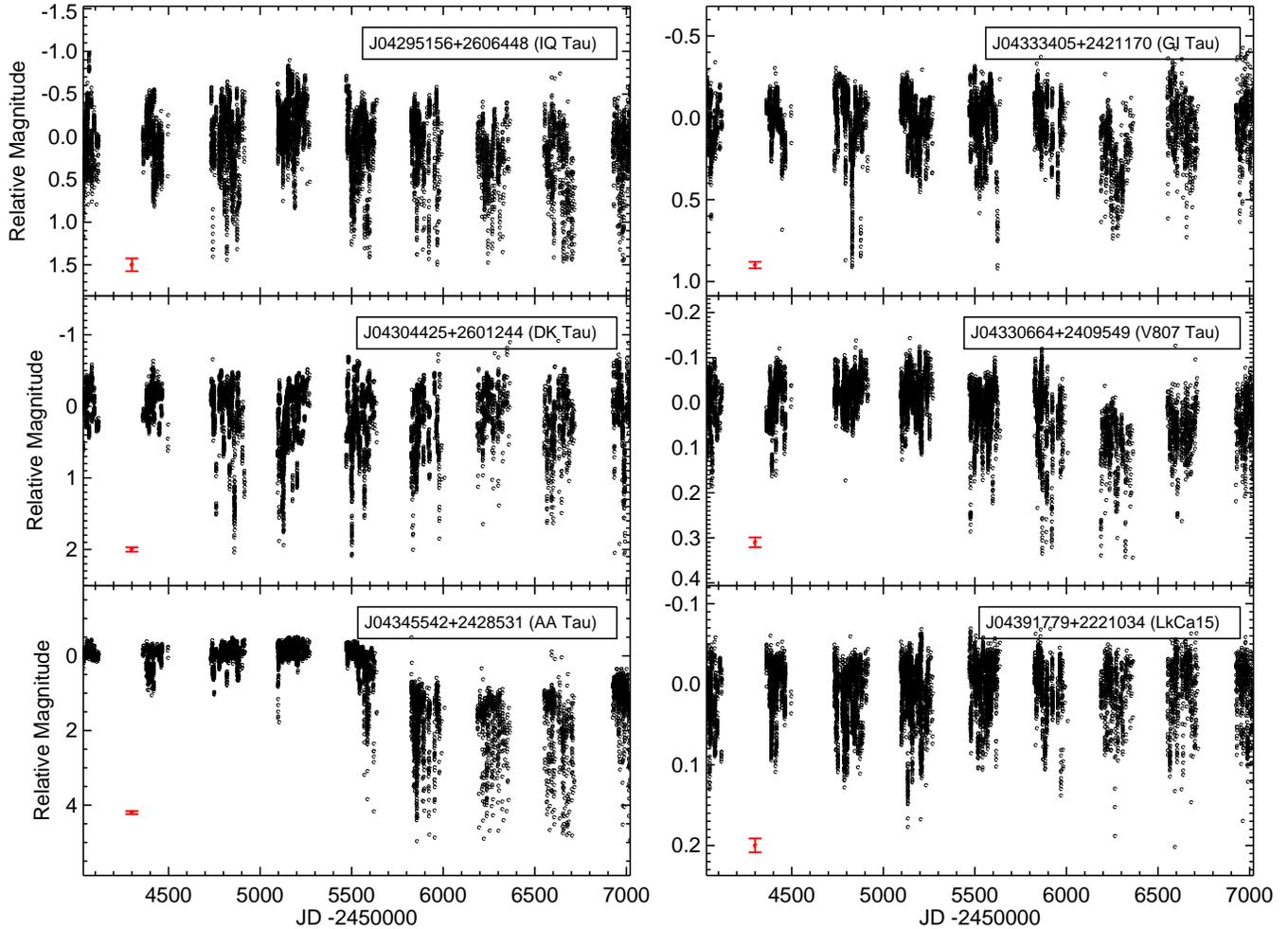}
  \caption{The KELT light curves for the six dipper stars identified in our sample. The median per-point error is shown on the bottom left of each plot in red.}
 \label{fig:Dippers}
\end{figure*}

\begin{figure*}[!ht]
  \centering
  \includegraphics[width=0.99\linewidth, trim = 0 2.3in 0 0]{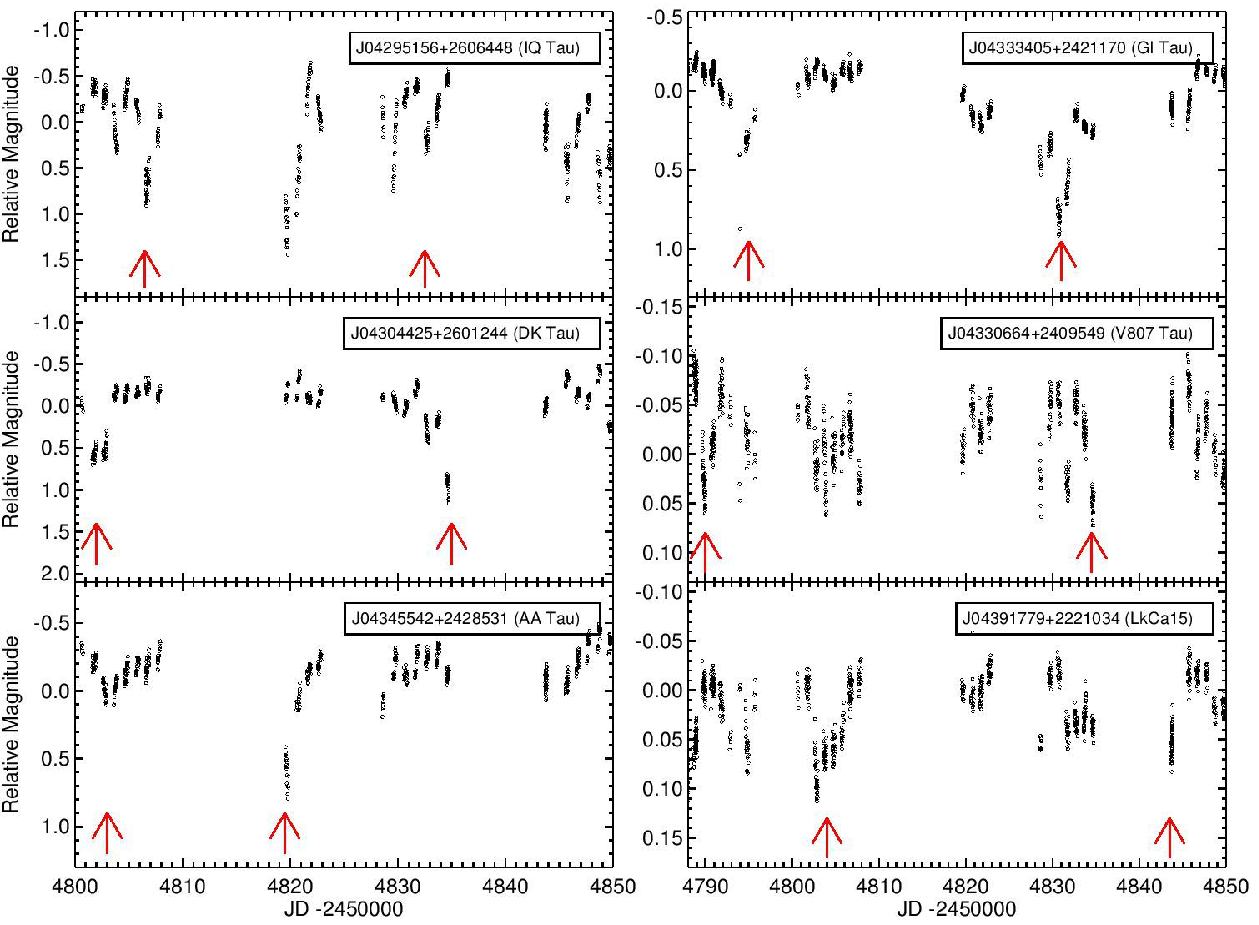}
  \caption{A zoom in on the six dippers identified showing the dipper phenomena at various depths. The arrows are marking some representative dips for each dipper, not necessarily all of them.}
 \label{fig:Dippers2}
\end{figure*}

\section{Identifying Variable and Periodic Objects}

\subsection{Variability Testing}
\label{sec:variability}

We employed the following four variability metrics to determine the variability of objects in the Taurus field, following the work of \citet{Wang:2013,Oelkers:2015}; and Oelkers et al. in prep (see Figure~\ref{fig:UXor}). These metrics help to identify large amplitude variability, which is not necessarily periodic. We outline each variability metric below but direct the reader to the previous works for a more detailed discussion of each metric. These metrics were determined empirically using all light curves in KELT-North field 03\footnote{We use field 03 because all but one star were found in 03 and we do not expect much field-to-field variation for each metric}. Any star which passed all 4 metric cutoffs was flagged as a variable star. Additionally, we flagged any star as a possible blend if another star has a magnitude within 1.5 of the target star's magnitude and is located within 2\arcmin ($\sim 5$~pix).

First we identify stars with light curves that show unusually high dispersion for their magnitude, using the \textit{rms} and $\Delta_{90}$ statistics. The \textit{rms} statistic identifies the magnitude range for $68\%$ of the data points in each light curve; the $\Delta_{90}$ statistics identifies the magnitude range for $90\%$ of the data points in each light curve. We compute the upper 2$\sigma$ envelopes of both statistics individually, as a function of magnitude, and assume an object is lying above these limits because of \textit{bona-fide} astrophysical variability. These statistics were not calculated with error weighting but because we wanted the envelopes to be based on stars with no apparent light curve variation, we applied an iterative 2.5$\sigma$ clipping to the sample, as a function of magnitude, prior to re-calculating the final 2$\sigma$ envelope of each metric.

Next we compute the Welch-Stetson \textit{J} and \textit{L} statistics \citep{Stetson:1996}. These two statistics are useful to compute the variability between subsequent data points on the sampling rate of the KELT survey, typically 10-30~min. These statistics are expected to produce a distribution of values centered at or near zero with a one-sided tail. Stars in this tail are expected to be variable. We remove objects with $J,L > 3$ and do a $2.5\sigma$ iterative clipping to determine the mean and standard deviation of the \textit{J} \& \textit{L} distributions prior to determining the cutoffs described below. This clipping allows us to calculate the distribution properties of \textit{J} \& \textit{L} using a population of stars which show minimal light curve variation. The Stetson \textit{J} \& \textit{L} values are shown in Table \ref{tab:VarPer}.

Finally, we applied a $+3\sigma$ cutoff of this tail in \textit{J} and the $+5\sigma_{J}$ cutoff of this tail in \textit{L} to select variable objects. We applied a $+5\sigma_{L}$ cutoff in \textit{L}, rather than $+3\sigma_{L}$ cutoff, because we found the larger limit helped to remove spurious objects which passed the first 3 metrics but showed features of known detector systematics while retaining objects that show variation consistent with astrophysical phenomena. 

\subsection{Periodicity Testing}
\label{sec:period}

Following the approach of \citet{Stassun:1999}, we also executed a search for periodic signals using a Lomb-Scargle (L-S) periodogram \citep{Lomb:1976, Scargle:1982}. We searched for periods between a minimum period of 0.5 days and a maximum period of 50 days using 2000 frequency steps. Additionally, we masked periods between 0.5 and 0.505~d and 0.97--1.04~d to avoid the most common detector aliases associated with the solar and sidereal day and selected the highest peak of the power spectrum as the candidate period. 

We then executed a boot-strap analysis, using 1000 Monte-Carlo iterations, where the dates of the observations were not changed but the magnitude values of the light curve were randomly scrambled \citep[see][]{Henderson:2012}. We recalculated the Lomb-Scargle power spectrum for each iteration and recorded the maximum peak power. If, after 1000 iterations, a maximum power of the boot-strap analysis was found to be larger than the power of the candidate period, the period was rejected as a false-positive.  

\begin{figure*}[!ht]
  \centering
  \includegraphics[width=0.99\linewidth, trim = 0 0in 0 0]{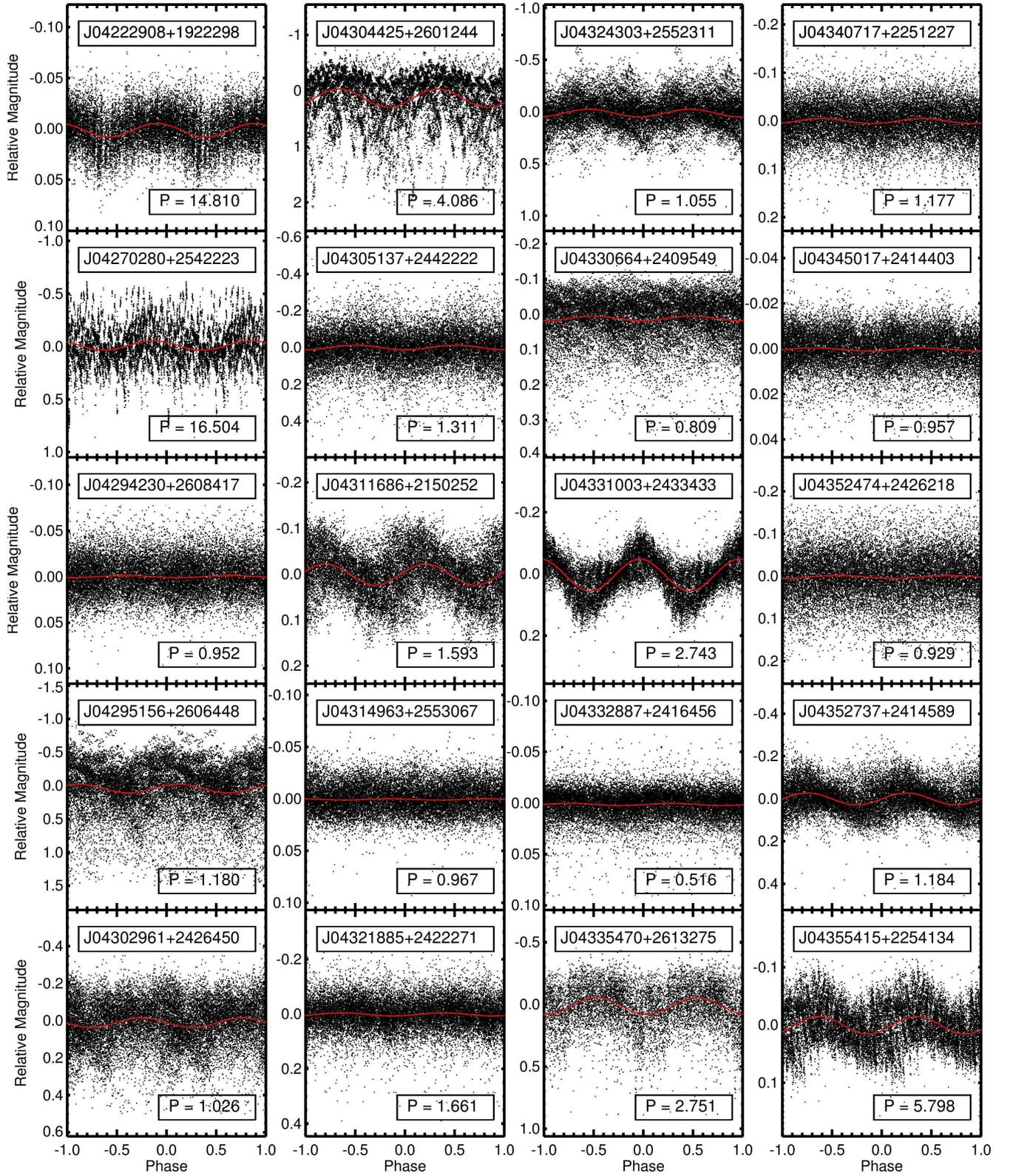}
  \caption{The KELT light curves for the identified periodic variables. The red line corresponds a sinusoidal fit to the phase-folded KELT observations.}
 \label{fig:Periodic}
\end{figure*}

\begin{figure*}[!ht]
  \centering
\includegraphics[width=0.99\linewidth, trim = 0 2.2in 0 0]{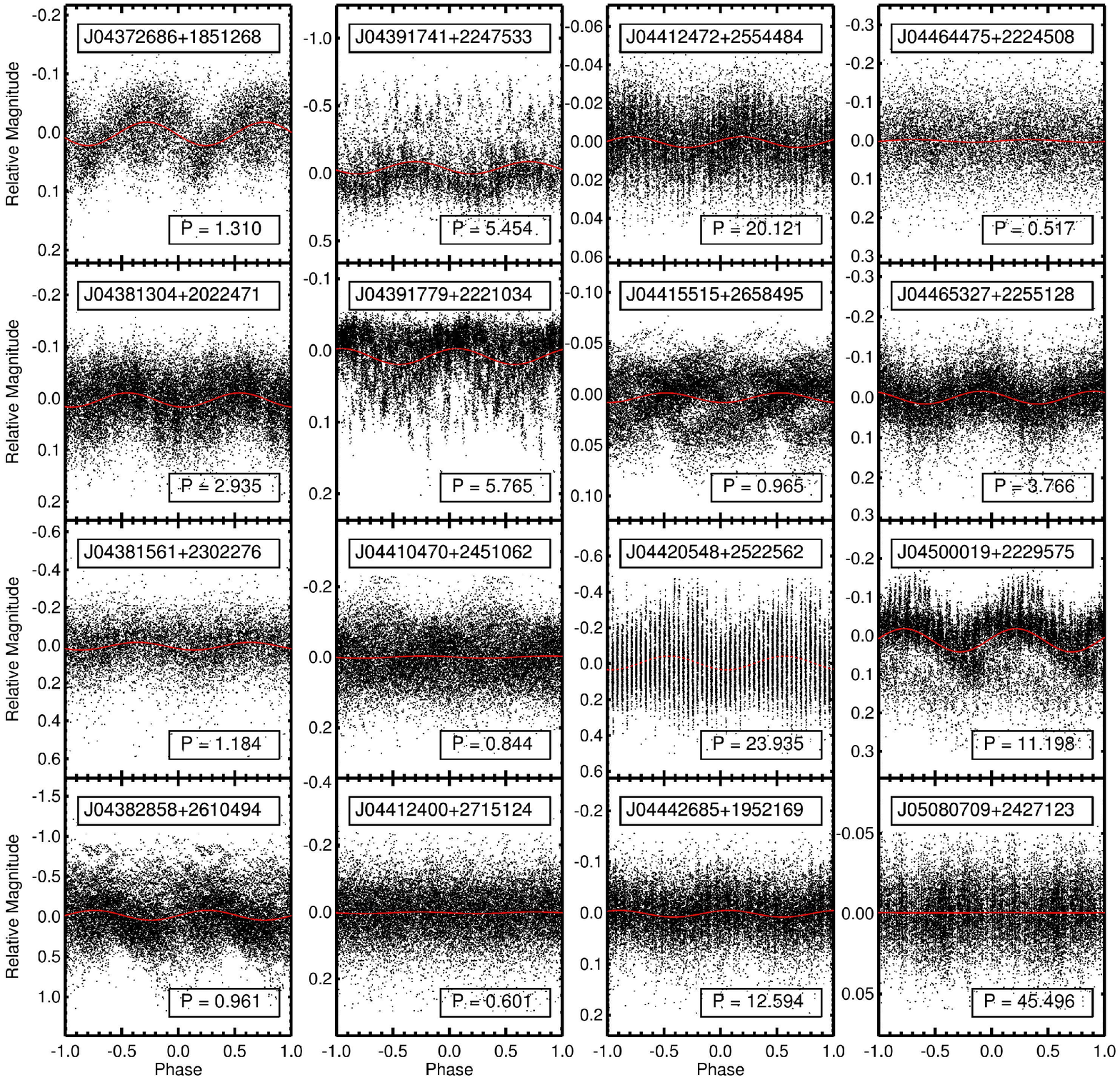}
  \caption{Figure 5 (continued): The KELT light curves for the identified periodic variables. The red line corresponds a sinusoidal fit to the phase-folded KELT observations.}
 \label{fig:Periodic2}
\end{figure*}

The top periodic signals found for each target using the technique described here are listed in Table \ref{tab:VarPer} and the corresponding phase folded light curve is shown in Figures \ref{fig:Periodic} and \ref{fig:Periodic2}. In general we interpret the periodic signals to likely be the rotation period of the primary star. A full interpretation of each individual case is beyond the scope of this paper.


\section{Interesting Variables}
In this section, we identify the interesting variable objects in our sample and summarize their properties based on the existing literature. This information will aid future follow-up observations of these targets and support the analysis of the \emph{K2} C13 photometry.

\subsection{Dippers and UXORs}
\label{sec:dippers}
Some YSOs, known as ``dippers", display short-duration large-amplitude dimmings that may be related to planet formation \citep{Cody:2010,Cody:2014}. This sub-class of YSOs consists of T~Tauri stars with optical light curves that exhibit very deep ($\sim$10--60\% in flux) and short-duration ($\sim$0.5--2 day) dimming events that are consistent with large dusty structures orbiting in the inner disk and transiting our line-of-sight to the star \citep[e.g.][]{Morales:2011,Cody:2014,Ansdell:2016a}. The flux dips can occur either quasi-periodically or aperiodically; when the dips are quasi-periodic, their periods are often similar to the stellar rotation period, suggesting that the occulting material is co-rotating with the star \citep[e.g.][]{Ansdell:2016a, Bodman:2016}. Although it was initially thought that dipper stars host nearly edge-on disks \citep[e.g.,][]{McGinnis:2015}, the handful of dippers with outer disk inclinations measured directly from resolved images exhibit a range of orientations, including face-on, intermediate, and nearly edge-on inclinations \citep{Ansdell:2016b,Scaringi:2016}.  The disks around dipper stars are in the early stages of planet formation, and these systems provide a unique probe of inner disk conditions during planet formation.

Another type of YSO with possible planet formation implications is UX Orionis (UXOR) stars, intermediate-mass pre-main sequence (PMS) stars (typically Herbig Ae stars) that also exhibit very deep dimming events that last for many months \citep[see review in ][]{Waters:1998}. It is unclear whether dippers and UXORs are produced by the same physical mechanism(s); but both classes of objects are thought to be related to occultations of the star by dusty circumstellar material. However, in contrast to lower-mass dippers, the dimming events of UXORs typically occur aperiodically and last weeks to months \citep{Grinin:1991}. It has been proposed that the UXOR dimming could be caused by hydrodynamical fluctuations of the inner-disk rim \citep{Dullemond:2003}.  Time-series multi-pass-band photometry, spectroscopy, and polarimetry can provide information on the gas and dust components of occulting material.   For example, UXOR dimming is accompanied by an increase in polarization as well as a color reversal (i.e., initial reddening and subsequent blue-ing), consistent with extinction by dust. 

To classify dippers in our sample, we visually analyzed the KELT light curves for short duration ($<$2 days), deep ($\sim$10--60\% in flux) dimming events. From this search, we identified six dippers and one UXOR star. The UXOR (HQ~Tau) and one of the dippers (AA~Tau) were previously identified in the literature. All of these sources have protoplanetary disks confirmed with sub-mm continuum observations. We summarize their properties below and provide their full light curves in Figures~\ref{fig:UXor} and \ref{fig:Dippers}. A zoom in of the dipper phenomena for the six identified dippers is shown in Figure \ref{fig:Dippers2}. Any dipper that also shows periodicity in their light curve is shown again in Figures \ref{fig:Periodic} and \ref{fig:Periodic2}. 
%EG: need a reference here? (MA: the references are many, and they are provided in the text below)

\subsubsection{AA~Tau (J04345542+2428531)}

AA~Tau is the prototypical dipper. Its photometric variability was originally characterized by \citet{Bouvier:1999} and the system has been extensively monitored in the several decades since its discovery \citep[e.g.][]{Bouvier:2003, Bouvier:2007, Bouvier:2013, Zhang:2015, Rodriguez:2015}. Apart from its dipper behavior, AA~Tau is a fairly typical T~Tauri star: it has a K7 spectral type, is a single star, and hosts a protoplanetary disk \cite[e.g.,][]{Bouvier:1999, Nguyen:2012, Andrews:2013}. The system has an accretion rate of log$\dot{M}_{\ast}=-8.48$ \msun yr$^{-1}$ and its protoplanetary disk has a dust mass of $M_{\rm dust}\approx30~M_{\oplus}$ \citep{Najita:2015}. The disk also has a high inclination and is seen nearly edge-on \citep{Menard:2003,Cox:2013}.

The optical light curve of AA~Tau shows a constant maximum brightness  $V\approx12.5$~mag, punctuated by regular dimming events every 8.2 days, where the dimming varied in depth but was typically $\sim$1.5~magnitudes. The dimming events have been interpreted as occultations of the central star by an inner disk warp located at a few stellar radii, where the inner disk warp is produced by interactions between an inclined stellar magnetosphere and a nearly edge-on accretion disk \citep{Bouvier:1999}. In 2011, AA~Tau suddenly dimmed by $\sim$2~mags in $V$-band and the regular flux dips were no longer detectable. The sudden enhanced extinction may be related to outer disk material coming into our line-of-sight as a result of Keplerian rotation or magnetic buoyancy instability \citep{Bouvier:2013, Rodriguez:2015, Zhang:2015}. 

The KELT data of AA~Tau have been previously published in \citet{Rodriguez:2015}, and we refer to this paper for a detailed analysis of the light curve. In short, the KELT data recovers the photometric characteristics of AA~Tau reported previously in the literature and summarized above, in particular its $\sim$8-day period and the sudden dimming in 2011. However, the reported $\sim$8-day periodicity is only recovered in individual KELT seasons, and not found when running the L-S periodogram on the entire nine seasons of KELT data, thus we do not report a period for AA Tau in Table \ref{tab:VarPer}. Note that the increase in photometric scatter seen in the KELT observations after the 2011 dimming of AA Tau (see Figure~\ref{fig:Dippers}) is due to the system being near the faintness limit of the telescope.

\subsubsection{LkCa~15 (J04391779+2221034)}

LkCa~15 is a single star of K5 spectral type with an accretion rate of log$\dot{M}_{\ast}=-8.87$ \msun yr$^{-1}$ \citep{Najita:2015}. It hosts a massive protoplanetary disk with $M_{\rm dust}\approx60 M_{\oplus}$ (\citealt{Najita:2015}; see also \citealt{Andrews:2013,Kraus:2012a}) at an intermediate inclination of $\sim$50$^{\circ}$ \cite[e.g.,][]{Andrews:2011,Thalmann:2014}. The disk is a so-called ``transitional disk" (TD) with a large inner dust gap extending to $\sim$50~AU at sub-mm wavelengths \citep{Andrews:2011}. \cite{Kraus:2012b} have also identified a possible forming protoplanet orbiting inside the dust gap using non-redundant aperture masking interferometry. However, recent observations of LkCa~15 with SPHERE have found no evidence of a protoplanet \citep{Thalmann:2016}.

The KELT-North data show a rotational signal at $P\approx5.7$~days superimposed with aperiodic drops in flux of $\sim$0.15~mag (see Figure \ref{fig:Periodic}). Although the dust cavity seen in the sub-mm may seem to contradict the dipper behavior, LkCa~15 is known to host an inner disk that could be providing the occulting dust. Its strong NIR excess, similar to the median SED of Taurus disks, indicates the presence of a compact and optically thick inner disk component near the dust sublimation radius \citep{Espaillat:2010}. Additionally, VLT/SPHERE images have resolved disk material in scattered light from 7--30 AU, which likely represents the optically thin regions of the inner disk component \citep{Thalmann:2015}.

\subsubsection{IQ~Tau (J04295156+2606448)}

IQ~Tau has an M0.5 spectral type and exhibits an accretion rate of log$\dot{M}_{\ast}=-7.55$ \msun yr$^{-1}$ \citep{Najita:2015}. The source hosts a protoplanetary disk with $M_{\rm dust}\approx30 M_{\oplus}$ \citep{Andrews:2013, Williams:2014,Najita:2015} and appears to be a single star based on high-resolution near-IR imaging \citep{Daemgen:2015} and high-resolution spectroscopy \citep{Nguyen:2012}. The inclination of IQ~Tau's disk has been estimated from sub-mm continuum and line emission to be $\sim$58$^{\circ}$ \citep{Guilloteau:2014}.

The KELT-North data show a periodic signal of $P\approx1.2$~days superimposed with aperiodic drops in brightness of up to $\sim$1~mag (see Figure \ref{fig:Periodic}). This period is not consistent with the $P=6.25$~day signal found by \cite{Bouvier:1995}. 

\subsubsection{DK~Tau (J04304425+2601244)}

DK~Tau is a K8+M1 binary system with a separation of $\sim$350~AU \citep{Akenson:2014, Andrews:2013}. The primary hosts a disk with $M_{\rm dust}\approx12 M_{\oplus}$ at an inclination of $41\pm11^{\circ}$, while the secondary hosts a disk with $M_{\rm dust}\approx2.0 M_{\oplus}$ at an inclination of $70\pm13^{\circ}$ \citep{Akenson:2014}. The system is also clearly accreting with log$\dot{M}_{\ast}=-7.42$ \msun yr$^{-1}$ \citep{Najita:2015}. 

The KELT-North data show a periodic signal at $P\approx4.1$~days, superimposed with deep aperiodic dips up to $\sim$2~mag in depth (see Figure \ref{fig:Periodic}). \cite{Bouvier:1995} found a rotational period of 8.4~days, which we also recover but at lower signal-to-noise. DK~Tau has been previously identified as a UXOR candidate by \cite{Oudmaijer:2001}, who used optical broadband photo-polarimetry to identify sources with significant photometric variability as well as increased polarization during dimming. In particular, while observing DK~Tau over two days, they found a dimming of $V\sim1$~mag accompanied by a doubling of the polarization from 1 to 2\%. However, our more extensive KELT-North data show that the dimming events last only a few days, which is more akin to dipper rather than UXOR behavior (See Figures \ref{fig:Dippers} and \ref{fig:Dippers2}).

\subsubsection{GI~Tau (J04333405+2421170)}

GI~Tau is a K7 star that hosts a protoplanetary disk with a relatively low dust mass of $M_{\rm dust}\approx5~M_{\oplus}$ but a high accretion rate of log$\dot{M}_{\ast}=-7.69$ \msun yr$^{-1}$ \citep{Najita:2015}. The disk inclination has not yet been directly constrained (e.g., by resolved imaging). It appears to be a single star based on high-resolution near-IR imaging \citep{Daemgen:2015} and high-resolution spectroscopy \citep{Nguyen:2012}. 

The KELT-North light curve exhibits dimming events that can be up to $\sim$1~mag in depth (see Figure~\ref{fig:Dippers} and \ref{fig:Dippers2}). We do not find a significant period, however \citet{Bouvier:1995} found a rotation period of 7.2~days.

\subsubsection{V807~Tau (J04330664+2409549)}

V807~Tau is a well-characterized triple system \cite[see][and references therein]{Schaefer:2012}. The two wider components are separated by $\sim$300~mas, with the secondary being itself a binary with a $\sim$40~mas separation. \cite{Schaefer:2012} used spatially resolved high-resolution spectra to determine that the primary has a K7 spectral type while the secondary is an M2-M2 binary. Moreover, their SED analysis showed IR excess consistent with an accretion disk around only the primary component of the system; there are no disk signatures for the secondary component. \cite{Andrews:2013} measured a composite sub-mm continuum flux consistent with $M_{\rm dust}\approx4 M_{\oplus}$. 

The KELT-North light curve exhibits a period of $P\approx0.8$ days, superimposed with aperiodic $\sim$0.2 mag drops in brightness (see Figure \ref{fig:Periodic}). The components of the triple system are unresolved in the KELT photometry, however the dips are unlikely to be caused by the stellar companions.

\subsubsection{HQ~Tau (J04354733+2250216)}

HQ~Tau has a K2 spectral type and hosts a protoplanetary disk with a relatively low dust mass of $M_{\rm dust}\approx2 M_{\oplus}$ \citep{Andrews:2013}. The disk inclination has not yet been directly constrained (e.g., by resolved imaging). The source appears to be a single star based on high-resolution imaging \citep{Simon:1995} and high-resolution spectroscopy \citep{Nguyen:2012}. 

Using Super-WASP photometry, \cite{Norton:2007} recovered a periodic signal of $P=2.4546$~days for HQ~Tau, which we recover in the first and fourth KELT-North seasons. However, because we do not recover a period when running the L-S periodogram on the entire nine seasons of KELT data, we do not report a period for HQ~Tau in Table \ref{tab:VarPer}. The second, fifth, and seventh KELT-North seasons show extended dimming events characteristic of UXOR variables: the dimming events reach $\sim$1.5-mag depths and last for weeks to months (See Figure \ref{fig:UXor}). UXOR behavior for this star has been previously reported in \cite{Watson:2015}.

%\subsection{Strong Periodic}
%These sources show sinusoidal periodic signal expected from star spots, but with large amplitudes above typical young stars. These may be particularly active stars.
%\subsubsection{HD~284496}
%HD~284496 is a WTTS lacking IR excess emission \citep{Spangler:2001}. We detect a clear rotational signal at $P=2.7$ days, which is the same value found by \cite{Norton:2007} using Super-WASP photometry. The amplitude of this large at 0.6~mag consistently throughout the KELT observing period.
%\subsubsection{V830~Tau}
%\subsubsection{IW~Tau}
%Binary  \citep{Simon:1995}

\subsection{Long-term Dimming}
\label{sec:longterm}

Two sources in our sample exhibit long-term dimming events that last for years. It is unclear whether these dimming events are related to circumstellar material or stellar variability.

\subsubsection{V1334~Tau (J04445445+2717454)}

V1334~Tau has a reported K1 spectral type \citep{Wichmann:1996}, but is also a multiple system that includes a close binary at $\sim$0.1$\arcsec$ separation with $\Delta K\approx2$~mag \citep{Daemgen:2015}. The system has been classified as a WTTS based on its low H$\alpha$ emission \citep{Wichmann:1996} and lacks the IR excess that would be indicative of a protoplanetary disk. 

A full analysis of the KELT-North data for V1334 Tau has been presented in \citet{Rodriguez:2017}. In short, the KELT-North light curve shows a long-duration dimming event of $\sim$0.12 mag beginning in 2009, which to date has not ended. Additionally, a $\sim$0.32 day periodicity was observed in every KELT season. It remains unclear what mechanisms are causing the long-term dimming and $\sim$0.32 day periodicity; see \citet{Rodriguez:2017} for a discussion of possible interpretations.

\subsubsection{V1341~Tau (J04500019+2229575)}

\begin{figure}[!ht]
  \centering
  \includegraphics[width=0.99\linewidth, trim = 0 7in 0 0]{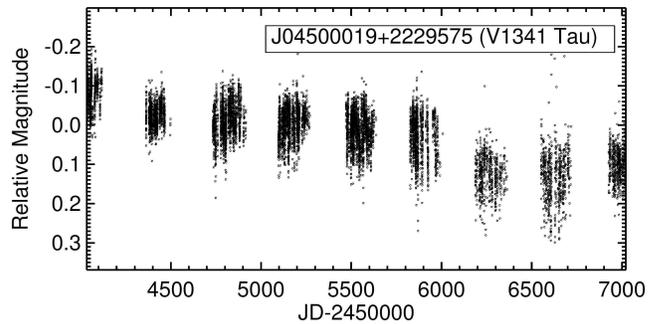}
  \caption{The KELT lightcurve for the long-term variable J04500019+2229575 (V1341 Tau).}
 \label{fig:LT}
\end{figure}

V1341~Tau has a K1 spectral type and has been classified as a WTTS based on its low H$\alpha$ emission \citep{Wichmann:1996}. It shows no evidence (e.g., IR excess) of a significant protoplanetary disk. V1341~Tau has candidate companions at 2.1$\arcsec$ and 8.4$\arcsec$ with $\Delta K$ values of 4.45 and 4.72, respectively \citep{Daemgen:2015}. The source gradually dims across all nine KELT-North seasons by $\sim$0.3 mags in total. We do not recover a significant periodic signal in any of the KELT seasons; the full light curve for this target is shown in Figure \ref{fig:LT}.

\subsection{Aperiodic Variability}

Here we present the targets in our sample that display photometric variability, often super-imposed on a periodic signal. These sources tend to have very high accretion rates, which is consistent with the observed photometric nature seen in their light curves. The phase-folded KELT light curves for sources with detected periodicity are shown in Figures \ref{fig:Periodic} and \ref{fig:Periodic2}. 

%EG: what is very high accretion rate in this context?

\subsubsection{DF~Tau (J04270280+2542223)}

DF~Tau is an M2.0-M2.5 binary at $99\pm14$~mas separation, with both components classified as CTTSs based on spectroscopic indicators \citep{Hartigan:2003}. The unresolved system is detected in the sub-mm continuum with a composite flux that corresponds to $M_{\rm dust}\approx2 M_{\oplus}$ in aggregate \citep{Andrews:2013}. Both components have particularly high accretion rates of log$\dot{M}_{\ast}=-6.9$ \msun yr$^{-1}$ \citep{Hartigan:2003}.

The KELT-North data show a periodicity of $P\approx16.5$~days. The periodic signal appears to be superimposed on accretion driven variability, as shown in Figure \ref{fig:Periodic}. Our measured period differs from that of \cite{Bouvier:1995}, who monitored DF~Tau with ground-based $UVBRI$ photometry and found a tentative period of 9.8~days with their $U$-band light curves only.

The nature of DF~Tau's photometric variability is consistent with the accretion state of this object. In particular, DF~Tau is thought to be in an unstable accretion regime based on the persistence of the redshifted absorption component in its H$\delta$ Balmer line \citep[see][and references therein]{Kurosawa:2013}. In this  regime, matter is thought to accrete along several transient accretion streams that appear in random locations, producing a aperiodic photometric light curve.

\subsubsection{CI~Tau (J04335200+2250301)}

CI~Tau has a K7 spectral type and hosts a massive protoplanetary disk with $M_{\rm dust}\approx60~M_{\oplus}$ \citep{Andrews:2013}. Resolved images reveal an extended disk with a weak non-axisymmetric feature \citep{Kwon:2015}. The system exhibits a notably high accretion rate of log$\dot{M}_{\ast}=-6.8$ \msun yr$^{-1}$ \citep{Hartigan:1995}, consistent with the photometric variability seen in its KELT-North light curve. 

The KELT-North data do not show a significant periodic signal. However, \citet{Johns-Krull:2016} found a tentative rotational period of $\sim$7.1~days based on 14 nights of ground-based $V$-band measurements. Variability in optical spectra has also been detected, which was interpreted as a temporary obscuration of a local hot region by circumstellar material \citep{Smith:1999}. 

This system was recently found to have a candidate young massive planet based on radial velocity (RV) variations in the optical and IR \citep{Johns-Krull:2016}. The RV amplitude yields $M \sin i \sim 8 M_{\rm Jup}$, which, in conjunction with a disk inclination of $\approx$46$^{\circ}$ estimated from sub-mm continuum emission \citep{Guilloteau:2014}, corresponds to a planet mass of $\sim$11--12~$M_{\rm Jup}$.

\subsubsection{DN~Tau (J04352737+2414589)}

DN~Tau is a CTTS with an M0 spectral type. It hosts a massive protoplanetary disk with $M_{\rm dust}\approx45~M_{\oplus}$ \citep{Andrews:2013} at an inclination of $\sim$30$^{\circ}$ \citep{Guilloteau:2014} based on sub-mm continuum and line images. DN~Tau has a fairly high accretion rate of log$\dot{M}_{\ast}=-8.45$ \msun yr$^{-1}$ \citep[e.g.,][]{Gullbring:1998}, although this is lower than the other sources in this sub-class.

We recover a period of $P\approx1.184$ days in the KELT-North data, with non-periodic variability superimposed over the periodic signal (see Figure \ref{fig:Periodic}). This period is not consistent with the previous estimate of a $\sim$6~day rotation period by \cite{Bouvier:1995}.

\subsubsection{DO~Tau (J04382858+2610494)}

DO~Tau is a CTTS with an M0 spectral type. It hosts a massive protoplanetary disk with $M_{\rm dust}\approx50~M_{\oplus}$ \citep{Andrews:2013}, and resolved images reveal a compact disk with an inclination of $\sim-32^{\circ}$ \citep{Kwon:2015}. The system has a notably high accretion rate of log$\dot{M}_{\ast}=-6.84$ \msun yr$^{-1}$ \citep{Gullbring:1998}, consistent with the variability seen in its KELT-North light curve. We recover a periodic signal of $\approx$0.96 daysin the KELT-North data. However \cite{Osterloh:1996} found a rotational period of $12.5\pm1.8$~days while monitoring the system over 10~days at optical wavelengths.

\subsubsection{V999~Tau (J04420548+2522562)}

V999~Tau is an M0.5-M2.5 binary separated by 0.27$\arcsec$. The secondary in V999~Tau has a fairly high accretion rate of log$\dot{M}_{\ast}=-8.24$ \msun yr$^{-1}$, while the primary has an upper limit of log$\dot{M}_{\ast}=-8.58$ \msun yr$^{-1}$ \citep{Hartigan:2003}. V1000 Tau (J04420732+2523032) and V995 Tau (J04420777+2523118) are both blended with the V999 Tau binary in the KELT-North data. However, we find a significant periodic signal of $P\approx23.9$~days, which appears to be superimposed with non-periodic variability, as shown in Figure \ref{fig:Periodic}.

\subsubsection{UZ~Tau (J04324303+2552311 \& J04324282+2552314)}
 
UZ~Tau is a quadruple system with UZ~Tau~Eab about 4$\arcsec$ away from UZ~Tau~Wab. UZ~Tau~E is an M1-M4 single-lined spectroscopic binary \citep{Mathieu:1996,Prato:2002}, while UZ~Tau~W is an M2-M3 sub-arcsecond binary separated by 0.34$\arcsec$ \citep{Hartigan:2003}. UZ~Tau~E has a very high accretion rate of log$\dot{M}_{\ast}=-5.7$ \citep{Hartigan:1995}, while both components in UZ~Tau~B also have high accretion rates of log$\dot{M}_{\ast}=-8.0$ \msun yr$^{-1}$ \citep{Hartigan:2003}, again consistent with the variability seen in their KELT-North lightcurve.

In the KELT-North data, we recover a periodic signal with $P\approx1.06$~days as shown in Figure \ref{fig:Periodic}. The recovered periodicity matches the period of the UZ~Tau~E spectroscopic binary. However, as this quadruple system is unresolved in the KELT data, the light curve should be interpreted with caution.

\section{Summary and Conclusion}
\label{sec:Conclusion}

In this paper, we have identified and visually classified the light curves of 56 YSOs in Taurus that were observed by the extended {\it Kepler} mission, {\it K2} during Campaign 13. Using observations from the KELT survey, we have identified six dippers (five previously unknown), a previously known UXor, six aperiodic variables, and two stars displaying long duration dimming events. Additionally, we use a L-S periodicity search on all the stars in our sample and identify any targets that display periodic variability. Table \ref{tab:CatalogueInfo2} reports the top L-S period recovered that is not a known alias and Figures \ref{fig:Periodic} and \ref{fig:Periodic2} show the phase-folded KELT light curves. 

KELT observations provide long-time baseline ($\le$10 years), high-cadence (10-30 min) photometry that is well-suited to studying both the short- and long-duration photometric variability of YSOs. The upcoming {\it Transiting Exoplanet Survey Satellite} ({\it TESS}, \citealp{Ricker:2014}), will obtain high-precision photometric observations of almost the entire sky for a duration of $\ge$27 days. Many young stellar associations are expected to be observed during the {\it TESS} mission. Similar to the work presented here, observations from ground-based surveys can complement the upcoming {\it TESS} campaigns by extending the observing time baseline by a decade or more. We encourage the community to use the identified variables from our analysis to obtain targeted simultaneous and/or complementary observations, such as multi-band photometry and time-series spectroscopy, to better understand the underlying processes causing the observed variability and their relation to planet formation. 

%EG: This next paragraph seems redundant; shorten or delete?
The data published with this paper are intended to provide a legacy data set to complement the very high-precision photometric observations of these targets by {\it K2}. We are providing the KELT light curves for known Taurus-Auriga stars in {\it K2} C13 (UT 2017 Mar 08	to UT 2017 May 27) with this paper. The KELT and {\it K2} photometry, combined with ground-based multi-band photometry and spectroscopy, will provide a comprehensive data set to study the underlying astrophysical processes related to star formation, stellar variability, and protoplanetary environments. 

\acknowledgments 
Early work on KELT-North was supported by NASA Grant NNG04GO70G. J.A.P. and K.G.S. acknowledge support from the Vanderbilt Office of the Provost through the Vanderbilt Initiative in Data-intensive Astrophysics. This work has made use of NASA's Astrophysics Data System and the SIMBAD database operated at CDS, Strasbourg, France. Work performed by J.E.R. was supported by the Harvard Future Faculty Leaders Postdoctoral fellowship. Work performed by P.A.C. was supported by NASA grant NNX13AI46G. Work by K.G.S. was supported by NSF PAARE grant AST-1358862. Work by D.J.S. and B.S.G. was partially supported by NSF CAREER Grant AST-1056524. G.S. acknowledges the support of the Vanderbilt Office of the Provost through the Vanderbilt Initiative in Data-intensive Astrophysics (VIDA) fellowship.

This research made use of Montage. It is funded by the National Science Foundation under Grant Number ACI-1440620, and was previously funded by the National Aeronautics and Space Administration's Earth Science Technology Office, Computation Technologies Project, under Cooperative Agreement Number NCC5-626 between NASA and the California Institute of Technology.

\bibliographystyle{apj}

\bibliography{Taurus_Dippers}

\begin{sidewaystable*}
\centering 
\scriptsize
\setlength\tabcolsep{1.5pt}
\caption{}
  \label{tab:CatalogueInfo}
  \begin{tabular}{ccccccccccccc}
  \hline
  \hline%
  2MASS ID & EPIC ID & RA J2000  &   DEC J2000 &   pmRA   &pmDE    &   $K_P$ mag  &   u$^{\prime}$ & g$^{\prime}$ & r$^{\prime}$ & i$^{\prime}$ & z$^{\prime}$ & Gaia Parallax$^1$\\
 &  & (Degrees)  &   (Degrees) &   (mas yr$^{-1}$)   &(mas yr$^{-1}$) &      &     &    &  & &  \\
     \hline
J04222908+1922298 & 210767482 & 65.62122028 & 19.37491611 & 10.50 & -37.30 & 10.68 & N/A $\pm$ N/A & N/A $\pm$ N/A & N/A $\pm$ N/A & N/A $\pm$ N/A & N/A $\pm$ N/A & 10.76 $\pm$ 0.227\\
J04270280+2542223 & 247986526 & 66.76162028 & 25.70621667 & 12.82 & -19.09 & 11.32 & 15.517 $\pm$ 0.017 & 11.888 $\pm$ 0.001 & 11.004 $\pm$ 0.001 & 10.344 $\pm$ 0.001 & 10.109 $\pm$ 0.001 & N/A $\pm$ N/A\\
J04274793+2430121 & 247813679 & 66.94972111 & 24.50334556 & 5.40 & -5.50 & 11.73 & 15.567 $\pm$ 0.008 & 15.176 $\pm$ 0.011 & 11.661 $\pm$ 0.001 & 11.192 $\pm$ 0.001 & 13.21 $\pm$ 0.013 & N/A $\pm$ N/A\\
J04294230+2608417 & 248045033 & 67.42626917 & 26.14492889 & 13.90 & -15.20 & 11.31 & 15.27 $\pm$ 0.008 & 12.236 $\pm$ 0.001 & 11.3 $\pm$ 0.001 & 10.912 $\pm$ 0.001 & 11.357 $\pm$ 0.002 & N/A $\pm$ N/A\\
J04295156+2606448 & 248040905 & 67.46482694 & 26.11245611 & 17.90 & -33.60 & 13.54 & 17.123 $\pm$ 0.01 & 15.622 $\pm$ 0.008 & 13.069 $\pm$ 0.002 & 12.13 $\pm$ 0.002 & 12.808 $\pm$ 0.013 & N/A $\pm$ N/A\\
J04302961+2426450 & 247805410 & 67.62340861 & 24.44584083 & 6.10 & -18.50 & 13.02 & 16.637 $\pm$ 0.007 & 15.332 $\pm$ 0.005 & 13.258 $\pm$ 0.003 & 12.081 $\pm$ 0.001 & 12.744 $\pm$ 0.011 & N/A $\pm$ N/A\\
J04303292+2602440 & 248032266 & 67.63719472 & 26.04559417 & 5.80 & -11.00 & 13.09 & 16.214 $\pm$ 0.007 & 16.287 $\pm$ 0.015 & 13.101 $\pm$ 0.001 & 12.75 $\pm$ 0.001 & 13.997 $\pm$ 0.018 & N/A $\pm$ N/A\\
J04304153+2430416 & 247814868 & 67.67311806 & 24.51152583 & 16.80 & -31.30 & 12.80 & 16.467 $\pm$ 0.008 & 16.171 $\pm$ 0.015 & 12.8 $\pm$ 0.001 & 12.13 $\pm$ 0.002 & 13.647 $\pm$ 0.013 & N/A $\pm$ N/A\\
J04304425+2601244 & 248029373 & 67.6843475 & 26.02351472 & 1.50 & -22.00 & 12.21 & 17.245 $\pm$ 0.027 & 14.056 $\pm$ 0.001 & 15.612 $\pm$ 0.012 & 14.669 $\pm$ 0.011 & 13.672 $\pm$ 0.013 & N/A $\pm$ N/A\\
J04305137+2442222 & 247843485 & 67.7141225 & 24.70620778 & 7.90 & -17.80 & 13.85 & 16.459 $\pm$ 0.007 & 15.511 $\pm$ 0.005 & 13.768 $\pm$ 0.003 & 12.205 $\pm$ 0.001 & 11.482 $\pm$ 0.003 & N/A $\pm$ N/A\\
J04305171+2441475 & 247842020 & 67.715473 & 24.696531 & 3.00 & -21.00 & 18.64 & N/A $\pm$ N/A & N/A $\pm$ N/A & N/A $\pm$ N/A & N/A $\pm$ N/A & N/A $\pm$ N/A & N/A $\pm$ N/A\\
J04311686+2150252 & 247454835 & 67.82024222 & 21.84035556 & 2.10 & -13.50 & 10.68 & N/A $\pm$ N/A & N/A $\pm$ N/A & N/A $\pm$ N/A & N/A $\pm$ N/A & N/A $\pm$ N/A & 8.038 $\pm$ 0.259\\
J04314963+2553067 & 248010721 & 67.95685056 & 25.88520778 & 7.63 & -40.31 & 9.73 & 11.246 $\pm$ 0.001 & 10.125 $\pm$ 0.001 & 9.632 $\pm$ 0.001 & 9.51 $\pm$ 0.001 & 10.536 $\pm$ 0.003 & 7.968 $\pm$ 0.362\\
J04321786+2422149 & 247794636 & 68.074424 & 24.370829 & 0.00 & -16.00 & 17.79 & N/A $\pm$ N/A & N/A $\pm$ N/A & N/A $\pm$ N/A & N/A $\pm$ N/A & N/A $\pm$ N/A & N/A $\pm$ N/A\\
J04321885+2422271 & 247795097 & 68.078635 & 24.37419667 & 6.60 & -19.70 & 13.45 & 18 $\pm$ 0.012 & 15.367 $\pm$ 0.004 & 14.772 $\pm$ 0.011 & 15.841 $\pm$ 0.014 & 12.619 $\pm$ 0.011 & N/A $\pm$ N/A\\
J04324282+2552314 & 248009353 & 68.17839361 & 25.87536611 & 0.00 & 0.00 & 12.06 & 16.423 $\pm$ 0.008 & 15.651 $\pm$ 0.006 & 12.383 $\pm$ 0.001 & 11.291 $\pm$ 0.002 & 11.505 $\pm$ 0.005 & N/A $\pm$ N/A\\
J04324303+2552311 & 248009353 & 68.17939917 & 25.87528556 & 0.00 & 0.00 & 12.06 & 16.423 $\pm$ 0.008 & 15.651 $\pm$ 0.006 & 12.383 $\pm$ 0.001 & 11.291 $\pm$ 0.002 & 11.505 $\pm$ 0.005 & N/A $\pm$ N/A\\
J04330664+2409549 & 247764745 & 68.27763417 & 24.16528306 & 10.50 & -20.40 & 10.95 & 15.164 $\pm$ 0.006 & 12.28 $\pm$ 0.001 & 12.253 $\pm$ 0.002 & 11.166 $\pm$ 0.001 & 10.532 $\pm$ 0.002 & N/A $\pm$ N/A\\
J04331003+2433433 & 247822311 & 68.29179167 & 24.56200889 & 3.60 & -21.40 & 11.93 & 15.9 $\pm$ 0.006 & 13.438 $\pm$ 0.002 & 11.652 $\pm$ 0.001 & 11.02 $\pm$ 0.001 & 10.977 $\pm$ 0.002 & N/A $\pm$ N/A\\
J04332887+2416456 & 247781229 & 68.37035361 & 24.27931194 & 23.37 & -17.71 & 7.87 & 9.569 $\pm$ 0.001 & 8.327 $\pm$ 0.001 & 7.753 $\pm$ 0.001 & 7.754 $\pm$ 0.001 & 9.149 $\pm$ 0.002 & N/A $\pm$ N/A\\
J04333405+2421170 & 247792225 & 68.39192028 & 24.35474806 & -3.00 & -15.60 & 13.03 & 24.355 $\pm$ 0.628 & 24.774 $\pm$ 0.511 & 22.489 $\pm$ 0.144 & 21.148 $\pm$ 0.066 & 21.801 $\pm$ 0.391 & N/A $\pm$ N/A\\
J04333456+2421058 & 247791801 & 68.39401333 & 24.35163028 & 0.00 & 0.00 & 12.72 & 16.537 $\pm$ 0.008 & 15.991 $\pm$ 0.013 & 12.357 $\pm$ 0.001 & 14.176 $\pm$ 0.012 & 12.957 $\pm$ 0.011 & N/A $\pm$ N/A\\
J04335200+2250301 & 247584113 & 68.4667325 & 22.84169556 & 11.90 & -15.40 & 12.57 & 15.808 $\pm$ 0.007 & 14.066 $\pm$ 0.002 & 12.404 $\pm$ 0.001 & 11.691 $\pm$ 0.001 & 13.036 $\pm$ 0.013 & N/A $\pm$ N/A\\
J04335470+2613275 & 248055184 & 68.47815389 & 26.22436333 & 34.20 & -11.30 & 13.87 & 18.587 $\pm$ 0.017 & 16.034 $\pm$ 0.005 & 16.589 $\pm$ 0.016 & 15.396 $\pm$ 0.014 & 14.337 $\pm$ 0.015 & N/A $\pm$ N/A\\
J04335562+2425016 & 247801362 & 68.48175194 & 24.41711667 & 3.90 & -4.10 & 13.41 & 22.072 $\pm$ 0.58 & 20.642 $\pm$ 0.095 & 19.202 $\pm$ 0.041 & 18.203 $\pm$ 0.028 & 17.09 $\pm$ 0.032 & N/A $\pm$ N/A\\
J04340717+2251227 & 247585953 & 68.52991 & 22.85627306 & 7.30 & -25.60 & 12.17 & N/A $\pm$ N/A & N/A $\pm$ N/A & N/A $\pm$ N/A & N/A $\pm$ N/A & N/A $\pm$ N/A & N/A $\pm$ N/A\\
J04342961+2423421 & 247798120 & 68.62335278 & 24.39502833 & -9.10 & -0.90 & 12.75 & 20.471 $\pm$ 0.102 & 18.999 $\pm$ 0.017 & 17.662 $\pm$ 0.008 & 17.009 $\pm$ 0.007 & 16.825 $\pm$ 0.019 & N/A $\pm$ N/A\\
J04345017+2414403 & 247776236 & 68.70904833 & 24.24452139 & 23.93 & -18.76 & 8.95 & 14.064 $\pm$ 0.009 & 9.735 $\pm$ 0.001 & 8.821 $\pm$ 0.001 & 8.558 $\pm$ 0.001 & 9.62 $\pm$ 0.003 & 5.299 $\pm$ 0.235\\
J04345542+2428531 & 247810494 & 68.73091556 & 24.48140444 & 4.30 & -15.30 & 13.72 & 15.86 $\pm$ 0.008 & 13.937 $\pm$ 0.002 & 12.206 $\pm$ 0.001 & 11.492 $\pm$ 0.001 & 11.307 $\pm$ 0.002 & N/A $\pm$ N/A\\
J04352089+2254242 & 247592507 & 68.83708944 & 22.90673722 & 10.10 & -14.50 & 13.30 & N/A $\pm$ N/A & N/A $\pm$ N/A & N/A $\pm$ N/A & N/A $\pm$ N/A & N/A $\pm$ N/A & N/A $\pm$ N/A\\
J04352474+2426218 & 247804500 & 68.853103 & 24.439415 & 12.00 & -36.00 & 17.06 & N/A $\pm$ N/A & N/A $\pm$ N/A & N/A $\pm$ N/A & N/A $\pm$ N/A & N/A $\pm$ N/A & N/A $\pm$ N/A\\
J04352737+2414589 & 247776947 & 68.86404417 & 24.24966056 & 4.50 & -21.70 & 11.86 & 15.613 $\pm$ 0.006 & 13.563 $\pm$ 0.002 & 11.903 $\pm$ 0.001 & 11.168 $\pm$ 0.001 & 11.081 $\pm$ 0.002 & N/A $\pm$ N/A\\
J04354733+2250216 & 247583818 & 68.94722333 & 22.83936806 & 11.80 & -13.20 & 11.86 & N/A $\pm$ N/A & N/A $\pm$ N/A & N/A $\pm$ N/A & N/A $\pm$ N/A & N/A $\pm$ N/A & N/A $\pm$ N/A\\
J04355277+2254231 & 247592463 & 68.96991611 & 22.90643861 & -4.50 & -11.20 & 13.41 & N/A $\pm$ N/A & N/A $\pm$ N/A & N/A $\pm$ N/A & N/A $\pm$ N/A & N/A $\pm$ N/A & N/A $\pm$ N/A\\
J04355349+2254089 & 247591948 & 68.97290556 & 22.90250056 & -10.60 & -24.80 & 13.88 & N/A $\pm$ N/A & N/A $\pm$ N/A & N/A $\pm$ N/A & N/A $\pm$ N/A & N/A $\pm$ N/A & N/A $\pm$ N/A\\
J04355415+2254134 & 247592103 & 68.97562917 & 22.90375806 & 10.80 & -15.40 & 10.78 & N/A $\pm$ N/A & N/A $\pm$ N/A & N/A $\pm$ N/A & N/A $\pm$ N/A & N/A $\pm$ N/A & 6.389 $\pm$ 0.269\\
J04372686+1851268 & 247119725 & 69.36196778 & 18.85741111 & 19.50 & -36.90 & 11.23 & 16.577 $\pm$ 0.009 & 15.851 $\pm$ 0.012 & 15.125 $\pm$ 0.011 & 14.293 $\pm$ 0.011 & 14.334 $\pm$ 0.016 & N/A $\pm$ N/A\\
J04381304+2022471 & 247280905 & 69.55436889 & 20.37972667 & 2.00 & -7.90 & 12.03 & 15.582 $\pm$ 0.007 & 12.935 $\pm$ 0.001 & 11.909 $\pm$ 0.001 & 11.513 $\pm$ 0.001 & 11.655 $\pm$ 0.003 & N/A $\pm$ N/A\\
J04381561+2302276 & 247609913 & 69.56504667 & 23.04099806 & -6.50 & -15.30 & 13.24 & 17.642 $\pm$ 0.012 & 15.588 $\pm$ 0.007 & 13.312 $\pm$ 0.002 & 12.5 $\pm$ 0.002 & 12.445 $\pm$ 0.01 & N/A $\pm$ N/A\\
J04382858+2610494 & 248049475 & 69.61910667 & 26.17988861 & 0.00 & 0.00 & 13.21 & 15.598 $\pm$ 0.008 & 15.697 $\pm$ 0.012 & 15.544 $\pm$ 0.013 & 14.482 $\pm$ 0.013 & 13.891 $\pm$ 0.018 & N/A $\pm$ N/A\\
J04391741+2247533 & 247578338 & 69.82258194 & 22.79815944 & 12.70 & -9.00 & 12.72 & N/A $\pm$ N/A & N/A $\pm$ N/A & N/A $\pm$ N/A & N/A $\pm$ N/A & N/A $\pm$ N/A & N/A $\pm$ N/A\\
J04391779+2221034 & 247520207 & 69.82414528 & 22.35094389 & 8.30 & -14.80 & 11.63 & 15.362 $\pm$ 0.007 & 15.464 $\pm$ 0.011 & 10.344 $\pm$ 0.001 & 13.74 $\pm$ 0.008 & 10.142 $\pm$ 0.001 & N/A $\pm$ N/A\\
J04410470+2451062 & 247864498 & 70.269635 & 24.85171056 & 4.80 & -19.90 & 12.23 & 16.265 $\pm$ 0.008 & 13.509 $\pm$ 0.001 & 12.014 $\pm$ 0.001 & 11.214 $\pm$ 0.001 & 10.994 $\pm$ 0.002 & N/A $\pm$ N/A\\
J04412400+2715124 & 248175684 & 70.34999111 & 27.25347556 & 3.60 & 5.00 & 12.87 & N/A $\pm$ N/A & N/A $\pm$ N/A & N/A $\pm$ N/A & N/A $\pm$ N/A & N/A $\pm$ N/A & N/A $\pm$ N/A\\
J04412472+2554484 & 248014510 & 70.35300833 & 25.91346333 & 1.30 & -2.10 & 10.24 & 15.693 $\pm$ 0.009 & 11.69 $\pm$ 0.001 & 10.168 $\pm$ 0.001 & 9.263 $\pm$ 0.001 & 9.651 $\pm$ 0.003 & 2.523 $\pm$ 0.406\\
J04415515+2658495 & 248145565 & 70.47980861 & 26.98039361 & -0.69 & -21.30 & 9.69 & N/A $\pm$ N/A & N/A $\pm$ N/A & N/A $\pm$ N/A & N/A $\pm$ N/A & N/A $\pm$ N/A & 9.314 $\pm$ 0.226\\
J04420548+2522562 & 247941378 & 70.52288944 & 25.38246222 & 2.90 & -10.90 & 14.17 & 18.277 $\pm$ 0.017 & 16.13 $\pm$ 0.004 & 14.35 $\pm$ 0.011 & 13.033 $\pm$ 0.001 & 11.946 $\pm$ 0.004 & N/A $\pm$ N/A\\
J04420732+2523032 & 247941613 & 70.53054389 & 25.38423111 & 1.00 & -28.10 & 14.57 & 18.478 $\pm$ 0.019 & 16.463 $\pm$ 0.004 & 14.363 $\pm$ 0.003 & 15.687 $\pm$ 0.014 & 11.669 $\pm$ 0.004 & N/A $\pm$ N/A\\
J04420777+2523118 & 247941930 & 70.53241139 & 25.38659222 & 5.30 & -9.20 & 14.31 & 18.298 $\pm$ 0.015 & 16.031 $\pm$ 0.004 & 14.119 $\pm$ 0.01 & 16.482 $\pm$ 0.015 & 11.935 $\pm$ 0.004 & N/A $\pm$ N/A\\
J04442685+1952169 & 247225984 & 71.11188194 & 19.87136611 & 36.90 & -38.90 & 12.08 & N/A $\pm$ N/A & N/A $\pm$ N/A & N/A $\pm$ N/A & N/A $\pm$ N/A & N/A $\pm$ N/A & N/A $\pm$ N/A\\
J04445445+2717454 & 248180268 & 71.22690167 & 27.29590389 & 7.20 & -22.60 & 9.46 & N/A $\pm$ N/A & N/A $\pm$ N/A & N/A $\pm$ N/A & N/A $\pm$ N/A & N/A $\pm$ N/A & N/A $\pm$ N/A\\
J04464475+2224508 & 247528573 & 71.68647833 & 22.41412833 & 0.20 & -3.50 & 12.75 & 16.321 $\pm$ 0.008 & 15.548 $\pm$ 0.012 & 12.819 $\pm$ 0.002 & 12.262 $\pm$ 0.001 & 12.354 $\pm$ 0.008 & N/A $\pm$ N/A\\
J04465327+2255128 & 247594260 & 71.7219625 & 22.92026278 & 33.20 & -15.80 & 12.44 & 16.545 $\pm$ 0.01 & 16.37 $\pm$ 0.019 & 12.397 $\pm$ 0.001 & 11.643 $\pm$ 0.001 & 11.749 $\pm$ 0.002 & N/A $\pm$ N/A\\
J04500019+2229575 & 247539775 & 72.50082778 & 22.4992925 & 0.60 & -17.60 & 11.17 & 15.296 $\pm$ 0.01 & 11.69 $\pm$ 0.001 & 10.959 $\pm$ 0.001 & 10.692 $\pm$ 0.001 & 10.897 $\pm$ 0.002 & 8.506 $\pm$ 0.248\\
J04504608+2126535 & 247406403 & 72.69201583 & 21.44821 & -3.00 & -4.30 & 11.50 & 15.963 $\pm$ 0.009 & 12.865 $\pm$ 0.001 & 11.502 $\pm$ 0.001 & 11.027 $\pm$ 0.001 & 11.351 $\pm$ 0.003 & N/A $\pm$ N/A\\
J05080709+2427123 & 247806504 & 77.02959806 & 24.45343444 & -11.00 & -15.90 & 15.35 & N/A $\pm$ N/A & N/A $\pm$ N/A & N/A $\pm$ N/A & N/A $\pm$ N/A & N/A $\pm$ N/A & N/A $\pm$ N/A\\
\hline
 \hline
\end{tabular}
\begin{flushleft}
  \footnotesize \textbf{\textsc{NOTES}} \\
  \footnotesize References:  u$^{\prime}$g$^{\prime}$r$^{\prime}$i$^{\prime}$z$^{\prime}$ are from \citet{Alam:2015}, $^1$\citet{Brown:2016} Gaia DR1 http: \\gea.esac.esa.int/archive/ 
%   \vspace{-.1in}
%  \footnotesize $^{a}$ The Multi-periodic classification indicates that multiple periods passed our cuts in the L-S analysis. 
%  \footnotesize $^{b}$ Only the top period that passed our cuts is shown. See \S \ref{sec:periodic} for any additional periods recovered. 
  \end{flushleft}
\end{sidewaystable*}

\begin{table*}
\centering 
\scriptsize
\setlength\tabcolsep{1.5pt}
\caption{}
  \label{tab:CatalogueInfo2}
  \begin{tabular}{cccccccccccccccc}
  \hline
  \hline%
2MASS ID & $J$ mag$^{1,2}$    &   $H$ mag$^{1,2}$    &   $K$ mag$^{1,2}$ & WISE 1$^{3}$ & WISE 2$^{3}$ & WISE 3$^{3}$ & WISE 4$^{3}$ \\ %Stetson J & Stetson H & Variable & Blend & Classification$^{a}$ & Period$^{b}$ \\
  \hline
J04222908+1922298 & 9.306 $\pm$ 0.02 & 8.894 $\pm$ 0.026 & 8.783 $\pm$ 0.017 & 8.729 $\pm$ 0.023 & 8.781 $\pm$ 0.019 & 8.714 $\pm$ 0.03 & 8.13 $\pm$ 0.277\\
J04270280+2542223 & 8.171 $\pm$ 0.026 & 7.256 $\pm$ 0.023 & 6.734 $\pm$ 0.024 & 6.028 $\pm$ 0.114 & 5.23 $\pm$ 0.068 & 3.829 $\pm$ 0.014 & 2.27 $\pm$ 0.024\\
J04274793+2430121 & 9.612 $\pm$ 0.022 & 9.273 $\pm$ 0.024 & 9.061 $\pm$ 0.018 & 8.928 $\pm$ 0.022 & 8.896 $\pm$ 0.021 & 8.92 $\pm$ 0.048 & 8.647 $\pm$ N/A\\
J04294230+2608417 & 9.594 $\pm$ 0.022 & 9.26 $\pm$ 0.019 & 9.106 $\pm$ 0.017 & 9.019 $\pm$ 0.023 & 9.013 $\pm$ 0.021 & 8.993 $\pm$ 0.044 & 8.098 $\pm$ N/A\\
J04295156+2606448 & 9.415 $\pm$ 0.02 & 8.417 $\pm$ 0.023 & 7.779 $\pm$ 0.023 & 7.27 $\pm$ 0.037 & 6.729 $\pm$ 0.023 & 4.906 $\pm$ 0.016 & 2.97 $\pm$ 0.016\\
J04302961+2426450 & 9.388 $\pm$ 0.024 & 8.398 $\pm$ 0.018 & 7.924 $\pm$ 0.016 & 7.4 $\pm$ 0.033 & 6.982 $\pm$ 0.02 & 5.133 $\pm$ 0.019 & 3.193 $\pm$ 0.029\\
J04303292+2602440 & 11.327 $\pm$ 0.021 & 10.92 $\pm$ 0.022 & 10.809 $\pm$ 0.023 & 10.722 $\pm$ 0.022 & 10.753 $\pm$ 0.021 & 10.814 $\pm$ 0.232 & 8.188 $\pm$ N/A\\
J04304153+2430416 & 10.216 $\pm$ 0.022 & 9.603 $\pm$ 0.022 & 9.375 $\pm$ 0.018 & 9.231 $\pm$ 0.023 & 9.197 $\pm$ 0.02 & 9.191 $\pm$ 0.044 & 8.89 $\pm$ N/A\\
J04304425+2601244 & 8.719 $\pm$ 0.03 & 7.758 $\pm$ 0.024 & 7.096 $\pm$ 0.016 & 6.12 $\pm$ 0.104 & 5.314 $\pm$ 0.055 & 3.599 $\pm$ 0.011 & 1.8 $\pm$ 0.012\\
J04305137+2442222 & 9.495 $\pm$ 0.021 & 8.695 $\pm$ 0.029 & 8.441 $\pm$ 0.021 & 8.106 $\pm$ 0.022 & 7.703 $\pm$ 0.021 & 5.96 $\pm$ 0.015 & 4.319 $\pm$ 0.024\\
J04305171+2441475 & 12.842 $\pm$ 0.023 & 11.435 $\pm$ 0.026 & 10.314 $\pm$ 0.023 & 8.56 $\pm$ 0.021 & 7.076 $\pm$ 0.021 & 4.449 $\pm$ 0.014 & 1.909 $\pm$ 0.016\\
J04311686+2150252 & 9.24 $\pm$ 0.034 & 8.83 $\pm$ 0.03 & 8.734 $\pm$ 0.02 & 8.673 $\pm$ 0.022 & 8.705 $\pm$ 0.019 & 8.631 $\pm$ 0.034 & 8.306 $\pm$ 0.376\\
J04314963+2553067 & 8.697 $\pm$ 0.02 & 8.431 $\pm$ 0.018 & 8.367 $\pm$ 0.027 & 8.323 $\pm$ 0.023 & 8.347 $\pm$ 0.02 & 8.32 $\pm$ 0.036 & 7.909 $\pm$ 0.324\\
J04321786+2422149 & 11.539 $\pm$ 0.02 & 10.791 $\pm$ 0.02 & 10.382 $\pm$ 0.02 & 10.092 $\pm$ 0.023 & 9.769 $\pm$ 0.02 & 9.556 $\pm$ 0.075 & 8.147 $\pm$ N/A\\
J04321885+2422271 & 9.538 $\pm$ 0.02 & 8.432 $\pm$ 0.021 & 8.106 $\pm$ 0.021 & 7.911 $\pm$ 0.026 & 7.829 $\pm$ 0.02 & 7.716 $\pm$ 0.023 & 7.659 $\pm$ 0.251\\
J04324282+2552314 & N/A $\pm$ N/A & N/A $\pm$ N/A & N/A $\pm$ N/A & N/A $\pm$ N/A & N/A $\pm$ N/A & N/A $\pm$ N/A & N/A $\pm$ N/A\\
J04324303+2552311 & 9.136 $\pm$ N/A & 8.117 $\pm$ N/A & 7.354 $\pm$ 0.033 & 6.252 $\pm$ 0.079 & 5.476 $\pm$ 0.045 & 3.69 $\pm$ 0.012 & 1.802 $\pm$ 0.016\\
J04330664+2409549 & 8.146 $\pm$ 0.023 & 7.357 $\pm$ 0.026 & 6.96 $\pm$ 0.016 & 6.473 $\pm$ 0.076 & 6.023 $\pm$ 0.026 & 5.181 $\pm$ 0.016 & 2.885 $\pm$ 0.021\\
J04331003+2433433 & 9.325 $\pm$ 0.021 & 8.613 $\pm$ 0.016 & 8.422 $\pm$ 0.021 & 8.412 $\pm$ 0.023 & 8.38 $\pm$ 0.02 & 8.264 $\pm$ 0.031 & 8.329 $\pm$ 0.368\\
J04332887+2416456 & 7.119 $\pm$ 0.018 & 7.006 $\pm$ 0.017 & 6.946 $\pm$ 0.017 & 6.907 $\pm$ 0.055 & 6.937 $\pm$ 0.021 & 6.999 $\pm$ 0.021 & 6.809 $\pm$ 0.117\\
J04333405+2421170 & 9.341 $\pm$ 0.02 & 8.418 $\pm$ 0.021 & 7.888 $\pm$ 0.023 & 7.087 $\pm$ 0.106 & 6.35 $\pm$ 0.072 & 4.011 $\pm$ 0.016 & 2.27 $\pm$ 0.027\\
J04333456+2421058 & 9.053 $\pm$ 0.027 & 8.108 $\pm$ 0.026 & 7.468 $\pm$ 0.021 & 6.568 $\pm$ 0.138 & 5.899 $\pm$ 0.093 & 3.84 $\pm$ 0.014 & 1.803 $\pm$ 0.016\\
J04335200+2250301 & 9.48 $\pm$ 0.02 & 8.431 $\pm$ 0.04 & 7.793 $\pm$ 0.02 & 6.756 $\pm$ 0.068 & 6.032 $\pm$ 0.038 & 4.367 $\pm$ 0.014 & 2.412 $\pm$ 0.026\\
J04335470+2613275 & 9.866 $\pm$ 0.025 & 8.591 $\pm$ 0.036 & 7.86 $\pm$ 0.026 & 7.396 $\pm$ 0.032 & 6.824 $\pm$ 0.02 & 5.252 $\pm$ 0.014 & 3.569 $\pm$ 0.025\\
J04335562+2425016 & 10.056 $\pm$ 0.02 & 9.286 $\pm$ 0.022 & 9.003 $\pm$ 0.017 & 8.854 $\pm$ 0.023 & 8.836 $\pm$ 0.02 & 8.789 $\pm$ 0.037 & 8.016 $\pm$ N/A\\
J04340717+2251227 & 9.606 $\pm$ 0.022 & 9.018 $\pm$ 0.04 & 8.755 $\pm$ 0.018 & 8.643 $\pm$ 0.023 & 8.624 $\pm$ 0.021 & 8.597 $\pm$ 0.03 & 8.38 $\pm$ N/A\\
J04342961+2423421 & 10.748 $\pm$ 0.022 & 10.339 $\pm$ 0.023 & 10.208 $\pm$ 0.018 & 10.095 $\pm$ 0.024 & 10.08 $\pm$ 0.021 & 10.143 $\pm$ 0.088 & 8.218 $\pm$ N/A\\
J04345017+2414403 & 7.425 $\pm$ 0.023 & 7.21 $\pm$ 0.02 & 7.111 $\pm$ 0.023 & 7.044 $\pm$ 0.051 & 7.023 $\pm$ 0.019 & 7.089 $\pm$ 0.019 & 6.981 $\pm$ 0.11\\
J04345542+2428531 & 9.433 $\pm$ 0.024 & 8.546 $\pm$ 0.023 & 8.047 $\pm$ 0.024 & 7.447 $\pm$ 0.036 & 6.76 $\pm$ 0.02 & 4.645 $\pm$ 0.014 & 2.505 $\pm$ 0.022\\
J04352089+2254242 & 9.781 $\pm$ 0.022 & 8.933 $\pm$ 0.024 & 8.592 $\pm$ 0.018 & 8.454 $\pm$ 0.023 & 8.403 $\pm$ 0.019 & 8.298 $\pm$ 0.026 & 8.453 $\pm$ 0.474\\
J04352474+2426218 & 14.576 $\pm$ 0.031 & 13.887 $\pm$ 0.037 & 13.788 $\pm$ 0.044 & 13.477 $\pm$ 0.028 & 13.352 $\pm$ 0.034 & 12.092 $\pm$ 0.439 & 8.966 $\pm$ 0.523\\
J04352737+2414589 & 9.139 $\pm$ 0.021 & 8.342 $\pm$ 0.027 & 8.015 $\pm$ 0.021 & 7.657 $\pm$ 0.029 & 7.216 $\pm$ 0.02 & 5.166 $\pm$ 0.016 & 3.039 $\pm$ 0.022\\
J04354733+2250216 & 8.655 $\pm$ 0.024 & 7.731 $\pm$ 0.016 & 7.135 $\pm$ 0.021 & 6.633 $\pm$ 0.071 & 5.415 $\pm$ 0.037 & 3.485 $\pm$ 0.015 & 1.51 $\pm$ 0.021\\
J04355277+2254231 & 9.549 $\pm$ 0.022 & 8.469 $\pm$ 0.065 & 7.625 $\pm$ 0.024 & 6.018 $\pm$ 0.057 & 5.733 $\pm$ 0.044 & 3.636 $\pm$ 0.016 & 1.431 $\pm$ 0.019\\
J04355349+2254089 & 10.041 $\pm$ 0.023 & 9.154 $\pm$ 0.022 & 8.797 $\pm$ 0.021 & 8.614 $\pm$ 0.033 & 8.547 $\pm$ 0.031 & 8.296 $\pm$ 0.032 & 8.189 $\pm$ N/A\\
J04355415+2254134 & 8.103 $\pm$ 0.02 & 7.489 $\pm$ 0.036 & 7.234 $\pm$ 0.016 & 7.118 $\pm$ 0.046 & 7.068 $\pm$ 0.026 & 6.849 $\pm$ 0.023 & 4.512 $\pm$ 0.046\\
J04372686+1851268 & 9.422 $\pm$ 0.041 & 8.558 $\pm$ 0.036 & 8.666 $\pm$ 0.054 & 8.691 $\pm$ 0.045 & 8.685 $\pm$ 0.044 & 8.543 $\pm$ 0.048 & 7.465 $\pm$ N/A\\
J04381304+2022471 & 10.074 $\pm$ 0.023 & 9.534 $\pm$ 0.023 & 9.358 $\pm$ 0.018 & 9.218 $\pm$ 0.022 & 9.262 $\pm$ 0.021 & 9.244 $\pm$ 0.039 & 8.15 $\pm$ N/A\\
J04381561+2302276 & 10.623 $\pm$ 0.023 & 9.931 $\pm$ 0.022 & 9.769 $\pm$ 0.021 & 9.661 $\pm$ 0.023 & 9.621 $\pm$ 0.02 & 9.551 $\pm$ 0.051 & 8.581 $\pm$ 0.389\\
J04382858+2610494 & 9.47 $\pm$ 0.022 & 8.243 $\pm$ 0.033 & 7.303 $\pm$ 0.017 & 6.341 $\pm$ 0.097 & 5.428 $\pm$ 0.053 & 3.48 $\pm$ 0.013 & 1.168 $\pm$ 0.014\\
J04391741+2247533 & 9.97 $\pm$ 0.023 & 9.26 $\pm$ 0.021 & 8.958 $\pm$ 0.02 & 8.591 $\pm$ 0.023 & 8.189 $\pm$ 0.021 & 6.227 $\pm$ 0.017 & 4.722 $\pm$ 0.028\\
J04391779+2221034 & 9.424 $\pm$ 0.02 & 8.6 $\pm$ 0.018 & 8.163 $\pm$ 0.018 & 7.504 $\pm$ 0.028 & 7.207 $\pm$ 0.019 & 5.696 $\pm$ 0.015 & 3.565 $\pm$ 0.022\\
J04410470+2451062 & 9.244 $\pm$ 0.022 & 8.479 $\pm$ 0.02 & 8.275 $\pm$ 0.026 & 8.213 $\pm$ 0.025 & 8.115 $\pm$ 0.02 & 8.012 $\pm$ 0.022 & 7.817 $\pm$ 0.22\\
J04412400+2715124 & 11.037 $\pm$ 0.022 & 10.568 $\pm$ 0.021 & 10.428 $\pm$ 0.02 & 10.351 $\pm$ 0.023 & 10.358 $\pm$ 0.02 & 10.375 $\pm$ 0.097 & 8.797 $\pm$ N/A\\
J04412472+2554484 & 6.994 $\pm$ 0.02 & 6.461 $\pm$ 0.033 & 6.169 $\pm$ 0.023 & 6.068 $\pm$ 0.114 & 5.849 $\pm$ 0.043 & 6.042 $\pm$ 0.016 & 5.88 $\pm$ 0.052\\
J04415515+2658495 & 8.365 $\pm$ 0.03 & 8.068 $\pm$ 0.017 & 7.973 $\pm$ 0.029 & 7.931 $\pm$ 0.026 & 7.952 $\pm$ 0.019 & 7.939 $\pm$ 0.02 & 7.782 $\pm$ 0.236\\
J04420548+2522562 & 9.787 $\pm$ 0.022 & 8.663 $\pm$ 0.034 & 8.227 $\pm$ 0.02 & 7.978 $\pm$ 0.024 & 7.776 $\pm$ 0.021 & 7.426 $\pm$ 0.018 & 6.218 $\pm$ 0.057\\
J04420732+2523032 & 9.58 $\pm$ 0.023 & 8.401 $\pm$ 0.024 & 7.945 $\pm$ 0.023 & 6.914 $\pm$ 0.019 & 5.378 $\pm$ 0.01 & 2.626 $\pm$ 0.006 & 3.046 $\pm$ 0.009\\
J04420777+2523118 & 9.811 $\pm$ 0.022 & 8.601 $\pm$ 0.021 & 7.942 $\pm$ 0.016 & 6.735 $\pm$ 0.034 & 6.322 $\pm$ 0.02 & 4.506 $\pm$ 0.013 & 2.747 $\pm$ 0.017\\
J04442685+1952169 & 9.563 $\pm$ 0.032 & 8.881 $\pm$ 0.04 & 8.707 $\pm$ 0.027 & 8.483 $\pm$ 0.022 & 8.464 $\pm$ 0.02 & 8.38 $\pm$ 0.024 & 8.145 $\pm$ 0.256\\
J04445445+2717454 & 7.734 $\pm$ 0.023 & 7.281 $\pm$ 0.036 & 7.154 $\pm$ 0.021 & 7.028 $\pm$ 0.048 & 7.046 $\pm$ 0.019 & 7.029 $\pm$ 0.019 & 6.878 $\pm$ 0.103\\
J04464475+2224508 & 10.52 $\pm$ 0.021 & 9.9 $\pm$ 0.022 & 9.728 $\pm$ 0.017 & 9.654 $\pm$ 0.023 & 9.704 $\pm$ 0.02 & 9.536 $\pm$ 0.046 & 8.395 $\pm$ N/A\\
J04465327+2255128 & 9.875 $\pm$ 0.022 & 9.234 $\pm$ 0.022 & 9.024 $\pm$ 0.018 & 8.942 $\pm$ 0.024 & 8.921 $\pm$ 0.021 & 8.838 $\pm$ 0.03 & 8.482 $\pm$ 0.486\\
J04500019+2229575 & 9.503 $\pm$ 0.024 & 9.021 $\pm$ 0.023 & 8.887 $\pm$ 0.018 & 8.819 $\pm$ 0.023 & 8.819 $\pm$ 0.02 & 8.729 $\pm$ 0.025 & 8.325 $\pm$ 0.276\\
J04504608+2126535 & 9.438 $\pm$ 0.022 & 8.861 $\pm$ 0.022 & 8.704 $\pm$ 0.017 & 8.634 $\pm$ 0.025 & 8.692 $\pm$ 0.023 & 8.601 $\pm$ 0.03 & 8.213 $\pm$ N/A\\
J05080709+2427123 & 11.396 $\pm$ 0.022 & 10.701 $\pm$ 0.022 & 10.481 $\pm$ 0.018 & 10.279 $\pm$ 0.023 & 10.104 $\pm$ 0.021 & 9.293 $\pm$ 0.036 & 5.561 $\pm$ 0.044\\\hline
\hline
\end{tabular}
\begin{flushleft}
  \footnotesize \textbf{\textsc{NOTES}} \\
  \footnotesize References: $^1$\citet{Cutri:2003}, $^2$\citet{Skrutskie:2006}, $^3$\citet{Cutri:2014}
%   \vspace{-.1in}
%  \footnotesize $^{a}$ The Multi-periodic classification indicates that multiple periods passed our cuts in the L-S analysis. 
%  \footnotesize $^{b}$ Only the top period that passed our cuts is shown. See \S \ref{sec:periodic} for any additional periods recovered. 
  \end{flushleft}
\end{table*}

\begin{table*}
\centering 
\scriptsize
\setlength\tabcolsep{1.5pt}
\caption{}
  \label{tab:VarPer}
  \begin{tabular}{ccccccccccc}
  \hline
  \hline%
&&&&&&&&\\
2MASS ID & Stetson J & Stetson H & RMS & $\Delta$90 &Variable$^{a}$ & Blend$^{b}$ & Classification$^{c}$ & Period & {$\dfrac{\rm Peak Power}{\rm Simulated Power}$}$^{d}$& Amplitude$^{e}$ \\
&&&&&&&&&&(Mag)\\
\hline
J04222908+1922298 & 0.355 & 0.207 & 0.018 & 0.058 & 0 & 0 & Periodic & 14.810 & 20.33 & 0.0052  \\
J04270280+2542223 & 7.349 & 5.063 & 0.187 & 0.605 & 1 & 0 & Periodic & 16.504 & 19.77 & 0.0500  \\
J04274793+2430121 & 0.046 & 0.016 & 0.020 & 0.065 & 0 & 0 & N/A & N/A & N/A & N/A  \\
J04294230+2608417 & 0.116 & 0.070 & 0.019 & 0.062 & 0 & 0 & Periodic & 0.952 & 8.90 & 0.0017  \\
J04295156+2606448 & 3.412 & 2.286 & 0.377 & 1.221 & 1 & 0 & Dipper/Periodic & 1.180 & 21.44 & 0.1010  \\
J04302961+2426450 & 1.148 & 0.774 & 0.123 & 0.399 & 1 & 0 & Periodic & 1.026 & 35.41 & 0.0297  \\
J04303292+2602440 & 0.258 & 0.129 & 0.108 & 0.350 & 0 & 0 & Blend & N/A & 6.42 & N/A  \\
J04304153+2430416 & 0.035 & 0.022 & 0.048 & 0.155 & 0 & 0 & N/A & N/A & N/A & N/A  \\
J04304425+2601244 & 10.724 & 7.233 & 0.420 & 1.319 & 1 & 0 & Dipper/Periodic & 4.086 & 33.10 & 0.1208  \\
J04305137+2442222 & 0.218 & 0.150 & 0.091 & 0.295 & 0 & 1 & Periodic & 1.311 & 14.16 & 0.0113  \\
J04305171+2441475 & 0.218 & 0.150 & 0.091 & 0.295 & 0 & 1 & Blend & N/A & N/A & N/A  \\
J04311686+2150252 & 2.193 & 1.467 & 0.050 & 0.167 & 1 & 0 & Periodic & 1.593 & 63.22 & 0.0249  \\
J04314963+2553067 & 0.974 & 0.350 & 0.013 & 0.044 & 0 & 0 & Periodic & 0.967 & 44.83 & 0.0026  \\
J04321786+2422149 & 0.138 & 0.088 & 0.053 & 0.175 & 0 & 1 & Blend & N/A & N/A & N/A  \\
J04321885+2422271 & 0.138 & 0.088 & 0.053 & 0.175 & 0 & 1 & Periodic & 1.661 & 4.15 & 0.0056  \\
J04324282+2552314 & 2.807 & 1.698 & 0.144 & 0.486 & 1 & 1 & Blend & N/A & N/A & N/A  \\
J04324303+2552311 & 2.807 & 1.698 & 0.144 & 0.486 & 1 & 1 & Periodic & 1.055 & 23.18 & 0.0440  \\
J04330664+2409549 & 3.862 & 2.588 & 0.062 & 0.198 & 1 & 0 & Dipper/Periodic & 0.809 & 44.71 & 0.0143  \\
J04331003+2433433 & 1.343 & 0.939 & 0.058 & 0.196 & 1 & 0 & Periodic & 2.743 & 196.47 & 0.0499  \\
J04332887+2416456 & 2.758 & 0.206 & 0.012 & 0.040 & 0 & 0 & Periodic & 0.516 & 12.11 & 0.0020  \\
J04333405+2421170 & 6.008 & 4.003 & 0.168 & 0.525 & 1 & 1 & Dipper & N/A & N/A & N/A  \\
J04333456+2421058 & 6.008 & 4.003 & 0.168 & 0.525 & 1 & 1 & Blend & N/A & N/A & N/A  \\
J04335200+2250301 & 1.612 & 1.101 & 0.126 & 0.423 & 1 & 0 & N/A & N/A & N/A & N/A  \\
J04335470+2613275 & 0.421 & 0.275 & 0.172 & 0.562 & 1 & 0 & Periodic & 2.751 & 16.98 & 0.0573  \\
J04335562+2425016 & 0.055 & 0.028 & 0.075 & 0.243 & 0 & 0 & N/A & N/A & N/A & N/A  \\
J04340717+2251227 & 0.094 & 0.059 & 0.036 & 0.118 & 0 & 0 & Periodic & 1.177 & 3.68 & 0.0044  \\
J04342961+2423421 & 0.030 & 0.016 & 0.068 & 0.221 & 0 & 0 & N/A & N/A & N/A & N/A  \\
J04345017+2414403 & 0.807 & 0.283 & 0.008 & 0.026 & 0 & 0 & Periodic & 0.957 & 20.20 & 0.0012  \\
J04345542+2428531 & 16.883 & 11.315 & 0.929 & 2.785 & 1 & 0 & Dipper/Periodic & 13.613 & 141.18 & 0.2628  \\
J04352089+2254242 & 0.034 & 0.022 & 0.124 & 0.397 & 0 & 0 & N/A & N/A & N/A & N/A  \\
J04352474+2426218 & 0.087 & 0.053 & 0.055 & 0.178 & 0 & 0 & Periodic & 0.929 & 5.78 & 0.0042  \\
J04352737+2414589 & 1.203 & 0.750 & 0.066 & 0.221 & 1 & 0 & Periodic & 1.184 & 43.23 & 0.0280  \\
J04354733+2250216 & 8.693 & 5.085 & 0.374 & 1.240 & 1 & 0 & UXor & N/A & N/A & N/A  \\
J04355277+2254231 & 2.155 & 1.391 & 0.035 & 0.117 & 1 & 1 & Blend & N/A & N/A & N/A  \\
J04355349+2254089 & 2.155 & 1.391 & 0.035 & 0.117 & 1 & 1 & Blend & N/A & N/A & N/A  \\
J04355415+2254134 & 2.155 & 1.391 & 0.035 & 0.117 & 1 & 1 & Periodic & 5.798 & 31.04 & 0.0130  \\
J04372686+1851268 & 1.047 & 0.713 & 0.043 & 0.139 & 1 & 0 & Periodic & 1.310 & 31.72 & 0.0212  \\
J04381304+2022471 & 0.413 & 0.238 & 0.048 & 0.155 & 0 & 0 & Periodic & 2.935 & 29.67 & 0.0109  \\
J04381561+2302276 & 0.099 & 0.061 & 0.108 & 0.353 & 0 & 0 & Periodic & 1.184 & 5.12 & 0.0155  \\
J04382858+2610494 & 3.219 & 2.268 & 0.306 & 0.994 & 1 & 0 & Periodic & 0.961 & 23.53 & 0.0707  \\
J04391741+2247533 & 1.303 & 0.896 & 0.192 & 0.654 & 1 & 0 & Periodic & 5.454 & 30.36 & 0.0510  \\
J04391779+2221034 & 2.794 & 1.877 & 0.035 & 0.114 & 1 & 0 & Periodic & 5.765 & 28.27 & 0.0118  \\
J04410470+2451062 & 0.983 & 0.695 & 0.069 & 0.224 & 1 & 0 & Periodic & 0.844 & 66.23 & 0.0339  \\
J04412400+2715124 & 0.084 & 0.044 & 0.072 & 0.232 & 0 & 0 & Periodic & 0.601 & 5.89 & 0.0063  \\
J04412472+2554484 & 0.718 & 0.343 & 0.014 & 0.046 & 0 & 0 & Periodic & 20.121 & 56.80 & 0.0025  \\
J04415515+2658495 & 2.990 & 1.933 & 0.025 & 0.081 & 1 & 0 & Periodic & 0.965 & 69.91 & 0.0067  \\
J04420548+2522562 & 1.511 & 1.041 & 0.135 & 0.443 & 1 & 1 & Periodic & 23.935 & 16.09 & 0.0301  \\
J04420732+2523032 & 1.511 & 1.041 & 0.135 & 0.443 & 1 & 1 & Periodic & 23.935 & 16.09 & 0.0301  \\
J04420777+2523118 & 1.511 & 1.041 & 0.135 & 0.443 & 1 & 1 & Periodic & 23.935 & 16.09 & 0.0301  \\
J04442685+1952169 & 0.093 & 0.059 & 0.043 & 0.142 & 0 & 0 & Periodic & 12.594 & 4.60 & 0.0057  \\
J04445445+2717454 & 5.744 & 4.258 & 0.037 & 0.122 & 1 & 0 & N/A & N/A & N/A & N/A  \\
J04464475+2224508 & 0.074 & 0.049 & 0.071 & 0.232 & 0 & 0 & Periodic & 0.517 & 4.65 & 0.0081  \\
J04465327+2255128 & 0.289 & 0.185 & 0.055 & 0.180 & 1 & 0 & Periodic & 3.766 & 25.13 & 0.0169  \\
J04500019+2229575 & 1.925 & 1.283 & 0.070 & 0.233 & 1 & 0 & Periodic & 11.198 & 48.06 & 0.0243  \\
J04504608+2126535 & 0.094 & 0.046 & 0.027 & 0.087 & 0 & 0 & N/A & N/A & N/A & N/A  \\
J05080709+2427123 & 0.068 & 0.044 & 0.016 & 0.054 & 0 & 0 & Periodic & 45.496 & 3.59 & 0.0014  \\
\hline
\hline
\end{tabular}
\begin{flushleft}
  \footnotesize \textbf{\textsc{NOTES}} \\
%   \vspace{-.1in}
  \footnotesize $^{a}$A star was flagged as a variable if it passed a 2$\sigma$ cut on the \textit{rms} and $\Delta_{90}$ statistic, a 3$\sigma$ cut on \textit{J}, and a 5$\sigma$ cut on \textit{L}. See section \ref{sec:variability} for a description of how the cuts were determined.  \\
  \footnotesize $^{b}$If another star is within 2\arcmin and 1.5 magnitude of the target, it is classified as a blend.\\
  \footnotesize $^{c}$Any target classified as ``N/A'' did not show a periodic signal or any variability that would classify it as a dipper or UXor. \\
  \footnotesize $^{d}$A measurement of how much larger the Lomb-Scargle peak of the actual period is versus the maximum peak of the 1000 simulated light curves. Any period where this ratio is $>$1 is defined as a real period. \\
\footnotesize $^{e}$For any target classified as ``N/A" the reported amplitude is the $\Delta_{90}$ statistic that identifies the magnitude range for $90\%$ of the data points. \\
\end{flushleft}
\end{table*}

%\begin{table*}
%\centering 
%\small
%\setlength\tabcolsep{1.5pt}
%\caption{}
%  \label{tab:StellarParam}
%  \begin{tabular}{ccccccc}
%  \hline
%  \hline%
%2MASS ID & Mass         & Rad.        & log(L)      & Age & Dist. & A$_{V}$\\
%         & M$_{\odot}$  & R$_{\odot}$ & L$_{\odot}$ & Myr & pc & mag \\
%\hline
%04222908+1922298 & 0.96$^{+0.50}_{-0.01}$ & 0.92$^{+0.47}_{-0.05}$ & 0.14$^{+0.42}_{-0.18}$ & 21$^{+38}_{-2}$ & 156$^{+23}_{-23}$ & 1.22$^{+0.73}_{-0.53}$ \\
%\hline
%\hline
%\end{tabular}
%\end{table*}

\end{document}